\documentclass[aps,11pt]{revtex4-1}
\usepackage{graphicx}
\usepackage{hyperref}
\usepackage{cancel}
\usepackage{amssymb}
\usepackage{textcomp}
\usepackage{amsmath}
\usepackage{bm}
\usepackage{times}
\usepackage{epsfig}
\usepackage{color}

\newcommand{\eV}{\textrm{eV}}
\newcommand{\GeV}{\textrm{GeV}}
\newcommand{\TeV}{\textrm{TeV}}
\newcommand{\fb}{\textrm{fb}}
\newcommand{\cm}{\textrm{cm}}

\begin{document}
 \title{\LARGE Radiative Linear Seesaw Model, Dark Matter and $U(1)_{B-L}$}
\bigskip

\author{Weijian Wang~$^{a}$}
\email{wjnwang96@aliyun.com}
\author{Zhi-Long Han~$^{b}$}
\email{hanzhilong@mail.nankai.edu.cn}

\affiliation{ $^a$~Department of Physics, North China Electric Power
University, Baoding 071003,China
\\
$^b$~School of Physics, Nankai University, Tianjin 300071, China}
\date{\today}

\begin{abstract}
In this paper we propose a radiated linear seesaw model where the
naturally small term $\mu_{L}$ are generated at one-loop level and
its soft-breaking of lepton number symmetry attributes to the
spontaneous breaking(SSB) of B-L gauge symmetry. The value of $B-L$
charges for new particles are assigned to satisfy the anomalies
cancelation. It is founded that some new particles may have exotic
values of $B-L$ charge such that there exists residual $Z_{2}\times
Z_{2}^{\prime}$ symmetry even after SSB of $B-L$ gauge symmetry. The
$Z_{2}\times Z_{2}^{\prime}$ discrete symmetry stabilizes the these
particles as dark matter candidates. In the model, two classes of
inert fermions and scalars with different $B-L$ charges are
introduced, leading to two-component dark matter candidates. The
lepton flavor violation processes, the relic density of dark matter,
the direct detection of dark matter and the phenomenology at LHC are
investigated.

\end{abstract}

\maketitle

\section{Introduction}
The origin of tiny but no-zero neutrino masses observed by neutrino
oscillation experiments\cite{neu1} remains so far a mystery, and so
provides us an opportunity to search the new physics beyond the
standard model(SM). Perhaps the simplest scenario which may
understand the neutrino puzzle is to introduce the Majorana mass
that breaks the global $B-L$ symmetry though the dimension-5
Weinberg operator $\lambda LL\Phi\Phi/\Lambda$\cite{wein}. This
effective operator can be realized though various pathways which
depends on the new physics scales the associated massages lie at.
For instance in the case of widely known type-I seesaw
mechanism\cite{type1} with right-handed Majorana ingredients $N_{R}$
as massages, one needs the super-heavy masses for $N_{R}$ i.e
$10^{14\sim16}$ GeV to fit the observed sub-eV neutrino mass. The
right-handed neutrinos are too heavy to be detected at future
experiments. In contrast, in the so-called low-scale scenarios, the
small neutrino mass is not only due to the massage state with heavy
mass but also to another naturally small mass parameter which breaks
the lepton number symmetry. This is the basic ideas behind many
schemes including type-II seesaw model\cite{typ2}, inverse seesaw
model\cite{inverse} and linear seesaw model\cite{linear, linear2}.
In these models, the mass of massage particles can be lowered down
to TeV or even hundreds GeV, a scale to be explored at collider
experiments.

On the other hand, the Planck data has shown that $26\%$ of the
energy density of our universe is occupied by dark matter. In the
view of particle physics, the weakly interacting massive
particles(WIMPs) are the most promising dark matter candidates. In
recent years, a class of models are proposed to incorporate the
neutrino mass puzzle and the existence of dark matter in a unified
framework. In these models, the neutrino masses are generated at
loop level and the dark matter is naturally contained as a inert
particle, where the $Z_{2}$ symmetry or a $U(1)_{X}$ symmetry is
used to guarantee the stability of dark matter. The radiated
generation of neutrino mass has been realized at one-loop
level\cite{1loop,Ma:2006km,1loope}, two-loop
level\cite{2loop,Ma:2007gq,Kanemura:2011mw,Kanemura:2014rpa,Aoki:2014cja} and
three-loop level\cite{3loop}. The systematic analysis of one and
two-loop realization for with possible topologies are performed in
Ref\cite{Bonnet:2012kz}.

In this paper, we proposed a radiated linear seesaw model where the
lepton number violation is due to the spontaneous symmetry
breaking(SSB) of $U(1)_{B-L}$ gauge symmetry, while the naturally
small mass parameter is generated at one-loop level. The linear
seesaw model was fist studied in the left-right theory with gauge
group $SU(3)\times SU(2)_{L}\times SU(2)_{R}\times
U(1)_{B-L}$\cite{linear}, and subsequently inspired on $SO(10)$
theory in the presence of gauge singlets\cite{linear2}. In linear
seesaw scenario, $\Psi_{R}$ and $\Psi_{L}$ are added onto SM so that
the Lagrangian is given by
\begin{equation}
L=M_{D}\bar{\nu}_{L}\Psi_{R}+M_{\Psi}\bar{\Psi}_{R}\Psi_{L}+\mu_{L}\tilde{\nu}_{L}\Psi_{L}+h.c
\end{equation}
The neutrino mass matrix in the basis of ($\nu_{L},
\Psi_{R}^{c},\Psi_{L})$ is
\begin{equation}
M_{\nu}=\left(\begin{array}{ccc}
0&M_{D}&\mu_{L}\\
M_{D}^{T}&0&M_{\Psi}\\
\mu_{L}^{T}&M_{\Psi}^{T}&0
\end{array}\right)
\end{equation}
The light neutrino mass is given by
$m_{\nu}\sim\mu_{L}M_{D}M_{\Psi}^{-1}$. Note that the $\mu_{L}$
violates the lepton number symmetry and plays the role of a
naturally small parameter. Thus it seems natural that there exists a
suppression mechanism where the $\mu_{L}$ term is generated via loop
diagram meanwhile the soft-breaking of lepton number symmetry may
attribute to the SSB of $B-L$ gauge symmetry. Moreover, when the
WIMPs as the dark matter candidates are involved in the loop
diagram, we can reasonably assume they are generated by a vacuum
expectation value with respect to the SSB of $B-L$ gauge symmetry at
TeV scale. These are the main motivations of this work.

Following the spirit of Ref.\cite{Kanemura:2014rpa}, the value of $B-L$ charges
should be carefully assigned since the anomalies cancelation must be
satisfied. It is founded that some new particles may have exotic
values of $B-L$ charge such that there exists residual $Z_{2}\times
Z_{2}^{\prime}$ symmetry even after SSB of $B-L$ gauge symmetry. The
$Z_{2}\times Z_{2}^{\prime}$ discrete symmetry stabilizes the these
particles from decaying to SM ingredients. Thus the lightest
particle with the same exotic value of $B-L$ charge can be a dark
matter candidate. In practise, we introduce two classes of inert
fermions and scalars to realize the model, leading to two component
dark matter candidates.

The existence of new fermions and scalars provides rich phenomenon.
Tiny neutrino masses are explained with one-loop induced
linear-seesaw-like mechanism. The charged-scalar mediates lepton
flavor violation (LFV) of charged leptons. The relic density and the
direct detection of the two component dark matter are investigated.
The properties of discovered SM Higgs will be changed by the new
particles. And these new particles provide plenty of new signatures
at LHC. Especially, multi-lepton signals with missing transverse
energy $\cancel{E}_T$ can be used to test our model. We find that
our model can satisfy current constraints from phenomenons mentioned
above.

The rest of paper is organised as follows. In Sec. \ref{model}, we
introduce the realization of radiative linear seesaw and
multi-component dark matter from gauged $U(1)_{B-L}$. In Sec.
\ref{phenom}, we discuss the phenomenon of lepton flavor violation,
dark matter and collider signatures. Conclusions are given in Sec.
\ref{conclu}.

\section{Model}
\label{model}
\subsection{Model Setup}

\begin{table}[!htbp]\large
\begin{tabular}{|c|c|c|c|c|c|c|c|c|c|c|}\hline\hline
Particles & $\Psi_{R}$ & $\Psi_{L}$ & $N_{R}$ & $N_{R}^{\prime}$ & $N_{R}^{\prime\prime}$& $\eta_{1}$ & $s_{1}$ & $\eta_{2}$ & $s_{2}$ & $\sigma$\\
\hline $SU(2)_{L}$ & \underline{1}
&\underline{1}&\underline{1}&\underline{1}&\underline{1}&\underline{2}&\underline{1}&\underline{2}&\underline{1}&\underline{1}\\
\hline$U(1)_{Y}$ & 0
&0&0&0&0&$\frac{1}{2}$&0&$\frac{1}{2}$&0&0\\
\hline$U(1)_{B-L}$ & -1
&0&$-\frac{1}{2}$&$x$&$-1-x$&$-\frac{1}{2}$&$-\frac{1}{2}$&$x$&$x$&1\\
\hline\hline
\end{tabular}
\caption{New particles content: $G_{SM}\times U(1)_{B-L}$}
\label{tabnew}
\end{table}

In our model the neutrino masses are generated via the diagram
depicted in Fig. \ref{Fig:mv}. The new particles content and their
charge assignment are listed in Table. \ref{tabnew}. We add $N_{\Psi}$
generation Weyl fermions $\Psi_{Ri}$, $\Psi_{Li}$, $N_{1}$
right-handed Majorana neutrinos $N_{R\alpha}$, and $N_{2}$ pairs of
right-handed Majorana neutrinos $N_{R}^{\prime}$ and
$N_{R}^{\prime\prime}$ to the SM where $i,\alpha$ and $\beta$ are
the generation indices. All the new fermions are singlets under SM
gauge group. Five new scalars $\eta_{1},s_{2},\eta_{2},s_{2}$ and
$\sigma$ are also added to SM. Because the new fermions are all SM
singlets, the $B-L$ gauge symmetry satisfies all anomaly
cancellations except for  $[U(1)_{B-L}]\times[Gravity]^{2}$ and
$[U(1)_{B-L}]^{3}$ \cite{Ma:2001kg}. Considering the conditions for
the absence of $[U(1)_{B-L}]\times[Gravity]^{2}$ and
$[U(1)_{B-L}]^{3}$ anomaly, one has
\begin{eqnarray}
3+(-\frac{1}{2})N_{1}+xN_{2}+(-1-x)N_{2}+(-1)N_{\psi}=0 \\
3+(-\frac{1}{2})^{3}N_{1}+x^{3}N_{2}+(-1-x)^{3}N_{2}+(-1)^{3}N_{\psi}=0
\end{eqnarray}
After solving the anomaly free condition, one obtains
\begin{equation}
N_{1}=2, \quad\quad N_{2}=1,\quad\quad N_{\psi}=1,\quad\quad
x=\frac{\sqrt{2}-1}{2}
\end{equation}
Thus we have the some inert particles classified into two parts. In
the first class, there are two Majorana right-handed neutrinos
$(N_{R1}$, $N_{R2})$ and the inert scalars $(\eta_{1},s_{1})$ with
their B-L charge being $-\frac{1}{2}$. In the second class, we
obtain a pair of Majorana right-handed neutrinos
$(N_{R}^{\prime},N_{R}^{\prime\prime})$ along with the inert scalars
$(\eta_{2},s_{2})$ whose the B-L charges are irrational numbers
($\frac{\sqrt{2}-1}{2}$ or $ \frac{-\sqrt{2}-1}{2}$). One notices
that the new particles with both $-\frac{1}{2}$ and the irrational
numbers can not decay into SM particles. Therefore the lightest
particles belonging to the same class is stable and can be regarded
as a dark matter candidate. The relevant Lagrangian for Yukawa
sector is given by
\begin{equation}\begin{split}\label{yukawa}
-L_{Y}=&y_{l}\overline{L_{l}}\psi_{R}i\tau_{2}\Phi^{\ast}+y^{\prime}\overline{\psi_{L}}\psi_{R}\sigma+h_{\alpha}\overline{N_{R\alpha}}\psi_{L}s_{1}
+f_{\alpha
l}\overline{L_{l}^{c}}N_{R\alpha}^{c}i\tau_{2}\eta_{1}^{\ast}+\frac{1}{2}Y_{\alpha}\overline{N_{R\alpha}^{c}}N_{R\alpha}\sigma\\
&+h\overline{N_{R}^{\prime}}\psi_{L}s_{2}+f_{
l}\overline{L_{l}^{c}}N_{R}^{\prime\prime
c}i\tau_{2}\eta_{2}^{\ast}+\frac{1}{2}Y\overline{N_{R}^{\prime\prime
c}}N_{R}^{\prime}\sigma+h.c
\end{split}\end{equation}
Without losing generality, we work in the basis where the mass term
of $N_{R1,2}$ is diagonal. As for the mass term of $N_{R}^{\prime}$
and $N_{R}^{\prime\prime}$, one can redefine the fields as
\begin{equation}
\chi_{1}=\frac{1}{\sqrt{2}}(N_{R}^{\prime}+N_{R}^{\prime\prime})\quad\quad\quad
\chi_{2}=\frac{i}{\sqrt{2}}(N_{R}^{\prime}-N_{R}^{\prime\prime})
\end{equation}
so that
\begin{equation}
\frac{1}{2}Y\overline{N_{R}^{\prime\prime
c}}N_{R}^{\prime}\sigma\rightarrow
\frac{1}{2}Y(\overline{\chi_{1}^{c}}\chi_{1}+\overline{\chi_{2}^{c}}\chi_{2})\sigma
\end{equation}
Now we get two Majorana neutrino eigenstates having the same masses.
Note that there is no interplay Yukawa terms between $N_{R1,2}$ and
$(N_{R}^{\prime}, N_{R}^{\prime\prime})$ because of the the B-L
charges assignment they have.

The scalar potential in our model is given by
\begin{equation}\begin{split}
V(\Phi,\sigma,\eta_{1},s_{1}, \eta_{2},s_{2})=&-\mu_{\Phi}^{2}\Phi^{\dag}\Phi+\lambda_{\Phi}(\Phi^{\dag}\Phi)^{2}-\mu_{\sigma}^{2}|\sigma|^{2}+\lambda_{\sigma}|\sigma|^{4}\\
&+\mu_{\eta_{1}}^{2}\eta_{1}^{\dag}\eta_{1}+\lambda_{\eta_{1}}(\eta_{1}^{\dag}\eta_{1})^{2}
+\mu_{\eta_{2}}^{2}\eta_{2}^{\dag}\eta_{2}+\lambda_{\eta_{2}}(\eta_{2}^{\dag}\eta_{2})^{2}\\
&+\mu_{s_{1}}^{2}|s_{1}|^{2}+\lambda_{s_{1}}|s_{1}|^{4}+\mu_{s_{2}}^{2}|s_{2}|^{2}+\lambda_{s_{2}}|s_{2}|^{4}+\lambda_{s_{1}s_{2}}|s_{1}|^{2}|s_{2}|^{2}\\
&+\lambda_{\eta_{1}\Phi}(\Phi^{\dag}\Phi)(\eta_{1}^{\dag}\eta_{1})+\lambda^{\prime}_{\eta_{1}\Phi}(\eta_{1}^{\dag}\Phi)(\Phi^{\dag}\eta_{1})
+\lambda_{\eta_{2}\Phi}(\Phi^{\dag}\Phi)(\eta_{2}^{\dag}\eta_{2})+\lambda^{\prime}_{\eta_{2}\Phi}(\eta_{2}^{\dag}\Phi)(\Phi^{\dag}\eta_{2})\\
&+\lambda_{\eta_{1}\eta_{2}}(\eta_{1}^{\dag}\eta_{1})(\eta_{2}^{\dag}\eta_{2})+\lambda_{\eta_{1}\eta_{2}}^{\prime}(\eta_{1}^{\dag}\eta_{2})(\eta_{2}^{\dag}\eta_{1})\\
&+\lambda_{s_{1}\Phi}|s_{1}|^{2}(\Phi^{\dag}\Phi)+\lambda_{s_{1}\eta_{1}}|s_{1}|^{2}(\eta_{1}^{\dag}\eta_{1})+\lambda_{s_{1}\eta_{2}}|s_{1}|^{2}(\eta_{2}^{\dag}\eta_{2})\\
&+\lambda_{s_{2}\Phi}|s_{2}|^{2}(\Phi^{\dag}\Phi)+\lambda_{s_{2}\eta_{1}}|s_{2}|^{2}(\eta_{1}^{\dag}\eta_{1})+\lambda_{s_{2}\eta_{2}}|s_{2}|^{2}(\eta_{2}^{\dag}\eta_{2})\\
&+\lambda_{\sigma\Phi}|\sigma|^{2}(\Phi^{\dag}\Phi)+\lambda_{\sigma\eta_{1}}|\sigma|^{2}(\eta_{1}^{\dag}\eta_{1})+\lambda_{\sigma\eta_{2}}|\sigma|^{2}(\eta_{2}^{\dag}\eta_{2})\\
&+\lambda_{s_{1}\sigma}|s_{1}|^{2}|\sigma|^{2}+\lambda_{s_{2}\sigma}|s_{2}|^{2}|\sigma|^{2}+(\mu_{1}s_{1}^{\dag}\Phi^{\dag}\eta_{1}+\mu_{2}s_{2}^{\dag}\Phi^{\dag}\eta_{2}+h.c)
\end{split}\end{equation}
where $\mu_{\Phi}^{2}$, $\mu_{\sigma}^{2}$, $\mu_{\eta_{1}}^2$,
$\mu_{\eta_{2}}^2$, $\mu_{s_{1}}^{2}$ and $\mu_{s_{2}}^{2}$ are
taken as positive values and the value coupling constants $\mu_{1}$
and $\mu_{2}$ in trilinear terms can be set as positive by
re-phasing $s_{1}$ and $s_{2}$. Notice that there is no terms like
$s_{1}\sigma^{2}$ or $s_{2}\sigma^{2}$ appearing in the scalar
potential. This has two fold meanings: First, the inert scalars
$\eta_{1,2}$ and $s_{1,2}$ do not acquire the VEV after the SSB of
$\Phi$ and $\sigma$; Second, there exists a residual $Z_{2}\times Z_{2}^{\prime}$
symmetry under which all the inert particles are odd even after the
breakdown of B-L symmetry. Therefore the residual $Z_{2}\times Z_{2}^{\prime}$ symmetry
stabilizes the inert particles, makes them to be two component dark matter
candidates.

\subsection{Matrices of Scalar Particles}
After the SSB, the scalar $\Phi$ and $\sigma$ is parameterized as
\begin{equation}
\Phi=\left(\begin{array}{c}
  G^{+}\\
  \frac{v_{\phi}+\phi_{0}+iG_{\phi}}{\sqrt{2}}
  \end{array}\right)\quad\quad\quad
  \sigma=\frac{v_{\sigma}+\sigma_{0}+iG_{\sigma}}{\sqrt{2}}
\end{equation}
where $v_{\phi}\simeq246$GeV is the VEV of the SM higgs doublet
scalar and $v_{\sigma}$ is responsible for the SSB of B-L symmetry
\cite{Khalil:2006yi}. The Nambu-Godstone bosons $G^{+}$, $G_{\phi}$
and $G_{\sigma}$ are absorbed by the longitudinal components of $W$,
 $Z$ and $Z^{\prime}$ gauge bosons. For simplicity, we ignore the
kinetic mixing between $U(1)_Y$ and $U(1)_{B-L}$ gauge boson
\cite{Langacker:2008yv}. Therefor the VEV $v_{\sigma}$ provides a
mass of $U(1)_{B-L}$ gauge boson $Z'$ as $M_{Z'}=g_{B-L}v_{\sigma}$,
where $g_{B-L}$ is the $U(1)_{B-L}$ gauge coupling constant. For the
extra gauge boson $Z'$, LEP-II provides a combined bound
$M_{Z'}/g_{B-L}>7~\TeV$ \cite{Cacciapaglia:2006pk}, which is just
the lower bound on $v_{\sigma}$. Then we obtain the mass matrix for
CP-even scalars $\phi_{0}$ and $ \sigma_{0}$
\begin{equation}
M^{2}(\phi_{0},\sigma_{0})=\left(\begin{array}{cc}
  \cos\theta&\sin\theta\\
  -\sin\theta&\cos\theta
  \end{array}\right)\left(\begin{array}{cc}
  M_{h}^{2}&0\\
  0&M_{H}^{2}
  \end{array}\right)\left(\begin{array}{cc}
  \cos\theta&-\sin\theta\\
  \sin\theta&\cos\theta
  \end{array}\right)
\end{equation}
where $h$ stands for the SM-like Higgs
\cite{Aad:2012tfa,Chatrchyan:2012ufa} and $H$ is an extra CP-even
Higgs boson \cite{Barger:2006sk,Barger:2007im,Robens:2015gla} with
the masses respectively as

\begin{eqnarray}
M_{h}^{2}=\lambda_{\Phi}v_{\phi}^{2}+\lambda_{\sigma}v_{\sigma}^{2}-\sqrt{(\lambda
v_{\phi}^{2}-\lambda_{\sigma}v_{\sigma}^{2})^{2}+\lambda_{\sigma\Phi}^{2}v_{\phi}^{2}v_{\sigma}^{2}}\\
M_{H}^{2}=\lambda_{\Phi}v_{\phi}^{2}+\lambda_{\sigma}v_{\sigma}^{2}+\sqrt{(\lambda
v_{\phi}^{2}-\lambda_{\sigma}v_{\sigma}^{2})^{2}+\lambda_{\sigma\Phi}^{2}v_{\phi}^{2}v_{\sigma}^{2}}
\end{eqnarray}
and mixing angle $\theta$ determined as
\begin{equation}
\sin 2\theta=\frac{2\lambda_{\sigma\Phi}
v_{\phi}v_{\sigma}}{M_{H}^{2}-M_{h}^{2}}
\end{equation}
On the other hand, the inert scalar $(\eta_{1}, s_{1})$ and
$(\eta_{2},s_{2})$ do not mix with $\Phi$ and $\sigma$ due to the
residual $Z_2$ symmetry. The mass matrix for inert scalar fields are
\begin{equation}
M(\eta_{1},s_{1},\eta_{2},s_{2})=(\eta_{1}^{\dag},s_{1}^{\dag},\eta_{2}^{\dag},s_{2}^{\dag})
\left(\begin{array}{cccc}
  M_{11}&M_{12}&0&0\\
  M_{21}&M_{22}&0&0\\
  0&0&M_{33}&M_{34}\\
  0&0&M_{43}&M_{44}
  \end{array}\right)
  \left(\begin{array}{c}
  \eta_{1}\\
  s_{1}\\
  \eta_{2}\\
  s_{2}
  \end{array}\right)
\end{equation}
where
\begin{equation}\begin{split}
&M_{11}=\mu_{\eta_{1}}^{2}+\frac{1}{2}\lambda_{\phi\eta_{1}}v_{\phi}^{2}+\frac{1}{2}\lambda_{\phi\eta_{1}}^{\prime}v_{\phi}^{2}
+\frac{1}{2}\lambda_{\sigma\eta_{1}}^{2}v_{\sigma}^{2}\\
&M_{22}=\mu_{s_{1}}^{2}+\frac{1}{2}\lambda_{s_{1}\phi}v_{\phi}^{2}+\frac{1}{2}\lambda_{s_{1}\sigma}v_{\sigma}^{2}\\
&M_{33}=\mu_{\eta_{2}}^{2}+\frac{1}{2}\lambda_{\phi\eta_{2}}v_{\phi}^{2}+\frac{1}{2}\lambda_{\phi\eta_{2}}^{\prime}v_{\phi}^{2}
+\frac{1}{2}\lambda_{\sigma\eta_{2}}^{2}v_{\sigma}^{2}\\
&M_{44}=\mu_{s_{2}}^{2}+\frac{1}{2}\lambda_{s_{2}\phi}v_{\phi}^{2}+\frac{1}{2}\lambda_{s_{2}\sigma}v_{\sigma}^{2}\\
&M_{12}=M_{21}=\frac{\mu_{1}}{\sqrt{2}}v_{\phi}\\
&M_{34}=M_{43}=\frac{\mu_{2}}{\sqrt{2}}v_{\phi}
\end{split}\end{equation}
There is also no mixing between $(\eta_{1},s_{1})$ and
$(\eta_{2},s_{2})$, therefor a residual $Z_2^{\prime}$ symmetry
between the two classes can be realised.
 After diagonalizing the mass matrix, we obtain
the mass eigenstates of inert scalars as
\begin{equation}
 \left(\begin{array}{c}
  A_{1,2}^{0}\\
  H_{1,2}^{0}
  \end{array}\right)=\left(\begin{array}{cc}
  \cos\theta_{1,2}&-\sin\theta_{1,2}\\
  \sin\theta_{1,2}&\cos\theta_{1,2}
  \end{array}\right)\left(\begin{array}{c}
  \eta_{1,2}^{0}\\
  s_{1,2}^{0}
  \end{array}\right)\quad,\quad \sin 2\theta_{1,2}=\frac{\sqrt{2}\mu_{1,2}v_{\phi}}{M_{A_{1,2}}^{2}-M_{H_{1,2}}^{2}}
\end{equation}
where
\begin{equation}
M_{A_{1,2}}^{2}=\frac{1}{2}\Big(M_{\eta_{1,2}}^{2}+M_{s_{1,2}}^{2}+\sqrt{(M_{\eta_{1,2}}^{2}-M_{s_{1,2}}^{2})^{2}+2\mu_{1,2}^{2}v_{\phi}^{2}}
\Big)
\end{equation}
\begin{equation}
M_{H_{1,2}}^{2}=\frac{1}{2}\Big(M_{\eta_{1,2}}^{2}+M_{s_{1,2}}^{2}-\sqrt{(M_{\eta_{1,2}}^{2}-M_{s_{1,2}}^{2})^{2}+2\mu_{1,2}^{2}v_{\phi}^{2}}
\Big)
\end{equation}
Here $M_{\eta_{1}}\equiv M_{11}$, $M_{s_{1}}\equiv M_{22}$,
$M_{\eta_{2}}\equiv M_{33}$ and $M_{s_{2}}\equiv M_{44}$.
\subsection{Neutrino Mass}\label{neutrinomass}

\begin{figure}[!htbp]
\begin{center}
\includegraphics[width=0.45\linewidth]{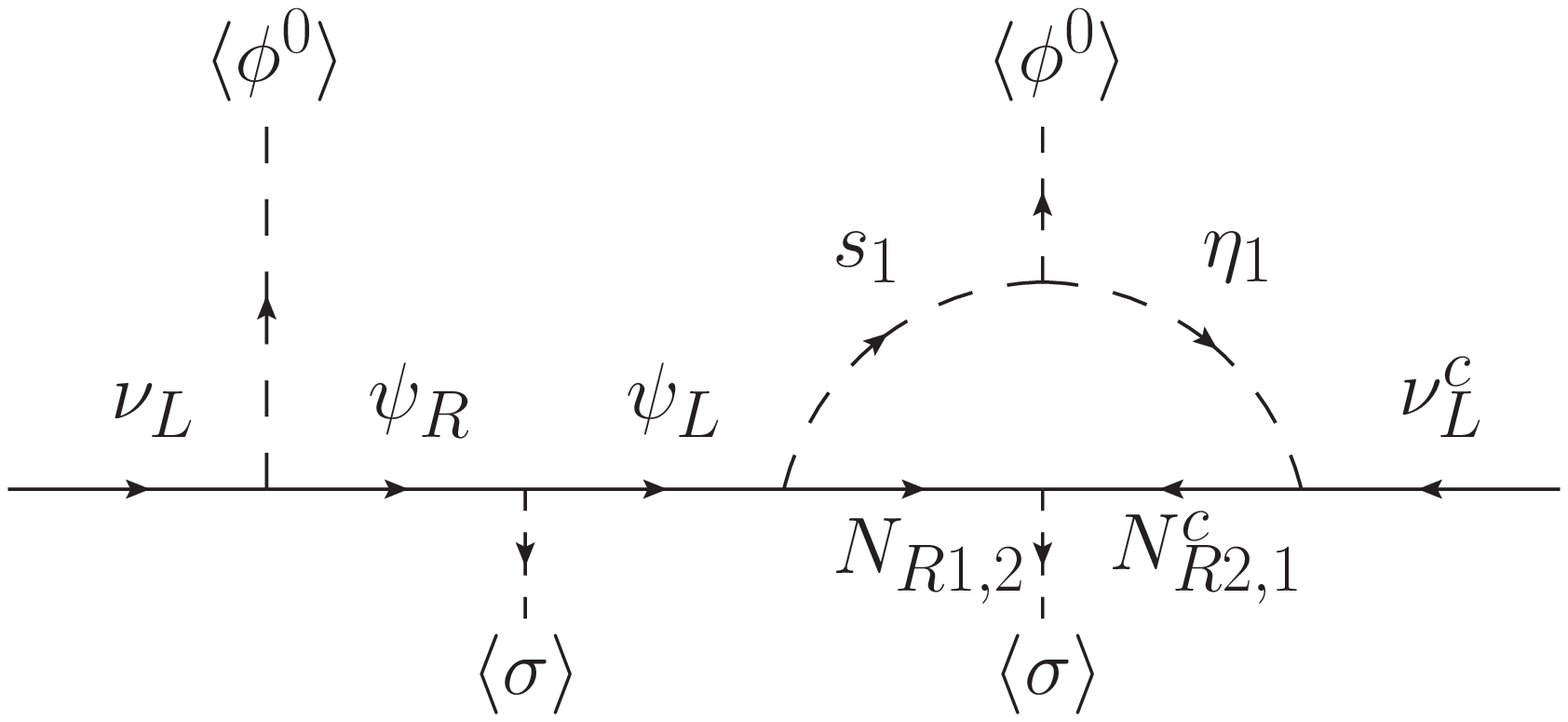}
\includegraphics[width=0.45\linewidth]{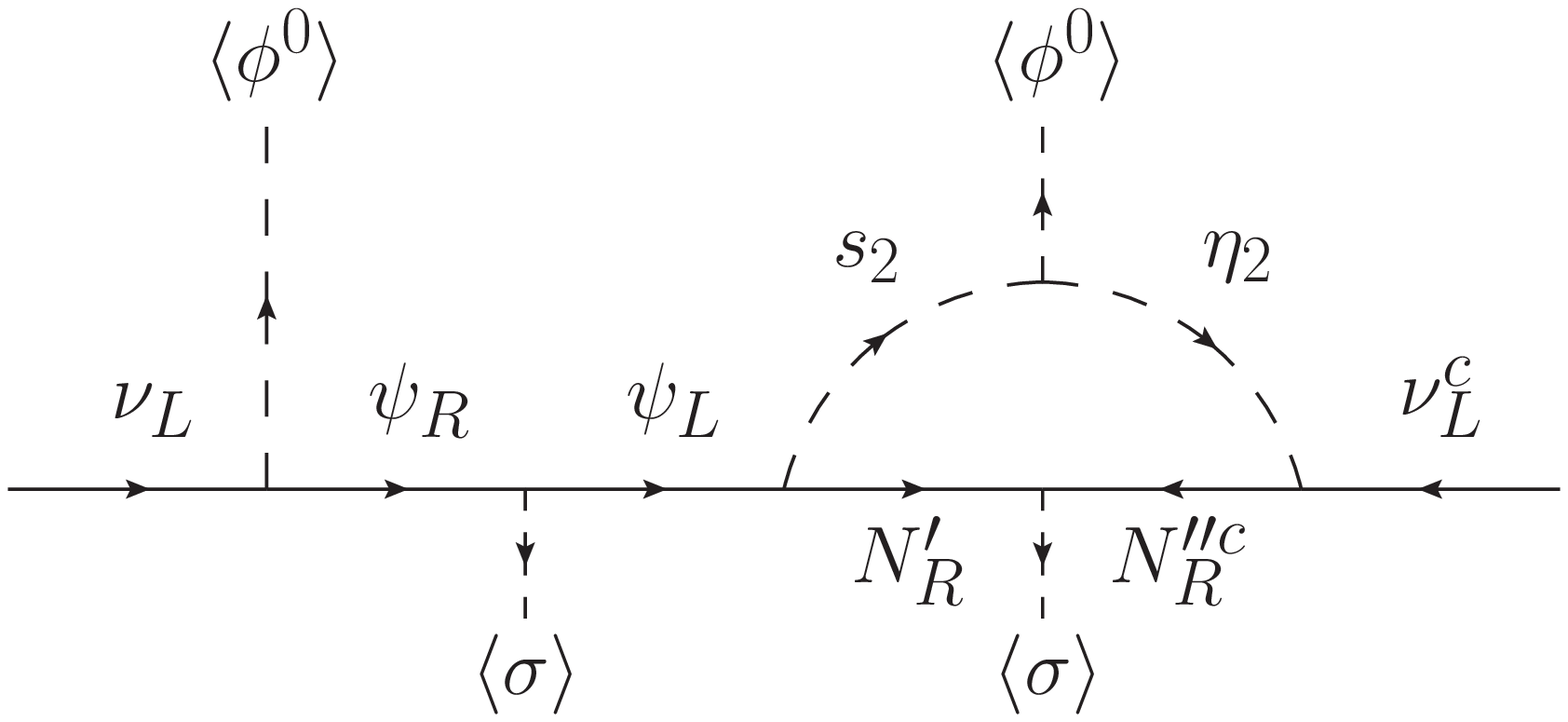}
\end{center}
\caption{The one-loop diagrams for neutrino masses in our model.}
\label{Fig:mv}
\end{figure}

As shown in Fig. \ref{Fig:mv}, the tiny neutrino mass are generated
by the linear seesaw mechanism except that the $\mu_{L}$ terms are
induced by a one-loop diagram. The effective mass matrix for active
neutrinos depicted in Fig. \ref{Fig:mv} is expressed as
\begin{equation}
M_{\nu ll^{\prime}}=M_{\nu ll^{\prime}}^{I}+M_{\nu ll^{\prime}}^{II}
\end{equation}
where
\begin{equation}\begin{split}
&M_{\nu
ll^{\prime}}^{I}=\frac{v_{\phi}\sin\theta_{1}\cos\theta_{1}}{16\pi^{2}\sqrt{2}M_{\psi}}y_{l}\sum_{i=1}^{2}
h_{i}f_{il^{\prime}}M_{i}\Big[\frac{M_{A_{1}^{2}}}{M_{i}^{2}-M_{A_{1}}^{2}}
\ln\big(\frac{M_{A_{1}}^{2}}{M_{i}^{2}}\big)-\frac{M_{H_{1}^{2}}}{M_{i}^{2}-M_{H_{1}}^{2}}\ln
\big(\frac{M_{H_{1}}^{2}}{M_{i}^{2}}\big)\Big]+(l\leftrightarrow l^{\prime})\\
&M_{\nu
ll^{\prime}}^{II}=\frac{v_{\phi}\sin\theta_{2}\cos\theta_{2}}{16\pi^{2}\sqrt{2}M_{\psi}}y_{l}hf_{l^{\prime}}
M_{\chi}\Big[\frac{M_{A_{2}^{2}}}{M_{\chi}^{2}-M_{A_{2}}^{2}}
\ln\big(\frac{M_{A_{2}}^{2}}{M_{\chi}^{2}}\big)-\frac{M_{H_{2}}^{2}}{M_{\chi}^{2}-M_{H_{2}}^{2}}\ln
\big(\frac{M_{H_{2}}^{2}}{M_{\chi}^{2}}\big)\Big]+(l\leftrightarrow
l^{\prime})
\end{split}\end{equation}
where $M_{i}(i=1,2)$ denotes the masses for $N_{R1}$ and $N_{R2}$;
$M_{\chi}$ denotes the masses for the eigenstates of
$N_{R}^{\prime}$ and $N_{R}^{\prime\prime}$. Tiny neutrino masses
can be obtained using the following benchmark points:
\begin{eqnarray}\label{bp}\nonumber
 \mu_1=\mu_2=0.1~\GeV,&&y=h=0.0028,~f=0.01, M_{\psi}=300~\GeV\\
 M_{N_{R1}}=149.5~\GeV,&&M_{N_{R2}}=200~\GeV,M_{\chi}=150~\GeV \\\nonumber
M_{A^0_1}=300~\GeV,&&M_{\eta^\pm_1}=270~\GeV,
M_{H^0_1}=1000~\GeV\\\nonumber
M_{A^0_2}=700~\GeV,&&M_{\eta^\pm_2}=690~\GeV, M_{H^0_2}=62~\GeV
\end{eqnarray}
with the index of Yukawa couplings suppressed for simplicity. Then
we get $M_\nu=0.0164~\eV(\sqrt{\Delta m_{13}^{2}})$.

The benchmark point given above seems rather unusual since the
$M_{\nu}$ becomes a rank-1 matrix as
\begin{equation}
\begin{split}
M_{\nu}\sim D\equiv\left(\begin{array}{ccc}
1&1&1\\
1&1&1\\
1&1&1
\end{array}\right)
\end{split}
\label{de}\end{equation} which is obviously not consistent with the
results of neutrino oscillation experiments. However, one reminds
the expression of matrix \eqref{de} is just the so-called flavor
democratic model studied by many authors and related to some flavor
symmetries\cite{zzx1}. The matrix $D$ can be diagonalized as
\begin{equation}
V_{\nu}^{T}DV_{\nu}=\left(\begin{array}{ccc}
0&0&0\\
0&0&0\\
0&0&3
\end{array}\right)\equiv \widehat{D}\quad\quad V_{\nu}=\frac{1}{\sqrt{6}}\left(\begin{array}{ccc}
\sqrt{3}&1&\sqrt{2}\\
-\sqrt{3}&1&\sqrt{2}\\
0&-2&\sqrt{2}
\end{array}\right)
\end{equation}
The unitary matrix $V_{\nu}$ corresponds to the democratic mixing
pattern. Since $\widehat{D}$ contains a dominant non-zero element at
$(3,3)$ position, the flavor democratic structure in $M_{\nu}$ in
eqn\eqref{de} can be viewed as a good approximation for our rank-2
neutrino mass matrix which exhibiting the strong and normal order of
neutrino mass spectrum. i.e $m_{1}=0\ll m_{2}=\sqrt{\Delta
m_{12}^{2}}\ll m_{3}=\sqrt{\Delta m_{13}^{2}}$. Thus the benchmark
point we take is reasonable. It is noted in the flavor basis the
democratic mixing matrix $V_{\nu}$ has already not been consistent
with the PMNS matrix $U_{PMNS}$ measured by experiments, however the
$V_{\nu}$ can be corrected by the charged lepton sector $V_{l}$ to
fit the neutrino oscillation data\cite{ccro}.

The small values of $\mu_{1,2}$ lead to
$\sin2\theta_{1,2}=(3.8,7.2)\times10^{-5}$, which plays a key rule
in the suppression of tiny neutrino masses. The choice of values of
$\mu_{1,2}$ is mainly for phenomenological consideration. First, in
case of scalar DM, we have $Z-S-S^*$ coupling proportional to
$\sin^2\theta_{1,2}$. The spin independent elastic cross section of
DM requires $\sin\theta_{1,2}<0.05$\cite{Akerib:2013tjd}, setting an
upper limit on $\mu_{1,2}\sim(10~\GeV)$ for electroweak(EW) scale
inert scalars. Second, in our benchmark point with inert particles
at the EW-scale and Yukawa couplings of the order $10^{-2}$ or
$10^{-3}$, rich phenomenon of new physics are expected for LHC and
LFV processes. On the other hand, larger values of $\mu_{1,2}$ is
possible if we decrease the Yukawa couplings to be more smaller
values. But this will predict a too small branch ratios for LFV
processes. Too small Yukawa coupling also seems unnatural from the
viewpoint of model building. Another solution is to increase the
mass of $\psi$ or inert particle to TeV-scale, which is beyond the
reach of LHC. For the DM candidate $N_{R1}$($H^0_2$), its mass is
set to be about half of the mass of the s-channel mediator $H$(h).
Then one obtains the large enough DM annihilation cross section to
account for the relic density. For the heavy dirac fermion $\psi$
and inert doublet scalars $\eta_1, A^1_0$, we choose their masses
around 300 GeV, so that they are testable at LHC. The other inert
scalars are around TeV-scale aiming to suppress the value of
neutrino mass.

\section{Phenomenology}
\label{phenom}

\subsection{Lepton Flavor Violation}

The Yukawa interactions of charged-scalar $\eta^{\pm}$ will
contribute to the LFV processes of charged leptons. Detail studies
on LFV processes in scotogenic models \cite{Ma:2006km} have been
carried out in Ref.\cite{Toma:2013zsa,Vicente:2014wga}. Currently,
the most severe constraint coming from MEG collaboration on muon
radiative decay with an upper limit BR($\mu\to e
\gamma$)$<5.7\times10^{-13}$ (90\% C.L.) \cite{Adam:2013mnn}.
In our model, the
analytical branching ratio of $\mu \to e \gamma$ is calculated as
\cite{Hisano:1995cp,Toma:2013zsa}:
\begin{equation}
\mbox{BR}(\mu\to e \gamma)=\frac{3\alpha_{em}}{64\pi
G^2_F}\left|\sum_{i=1}^{2}\frac{f_{i\mu}f_{ie}^*}{M_{\eta^+_1}^2}F\left(\frac{M_{N_i}^2}{M_{\eta^+_1}^2}\right)+
\frac{f_{l}f_{l}^*}{M_{\eta^+_2}^2}F\left(\frac{M_{N^{\chi}}^2}{M_{\eta^+_2}^2}\right)\right|^2,
\end{equation}
where the loop function $F(x)$ is:
\begin{equation}
F(x)=\frac{1-6x+3x^2+2x^3-6x^2\mbox{ln}x}{6(1-x)^4}.
\end{equation}
The benchmark point in Eqn.\ref{bp} predicts BR$(\mu\to e
\gamma)=9.8\times10^{-14}$, which satisfies the current limit and is
in the reach of future sensitivity \cite{Baldini:2013ke}. The limits
on $\tau$ observables are less stringent
\cite{Aubert:2009ag,Hayasaka:2010np}. With the much natural Yukawa
structure in our benchmark point, the predicted BR$(\tau\to \mu
\gamma)=5.8\times10^{-13}$ is far beyond the future sensitivity. But
on the other hand, hierarchal Yukawa structure
$|f_{ie}|\lesssim|f_{i\mu}|\lesssim|f_{i\tau}|$ with
$f_{i\tau}\sim\mathcal{O}(1)$ is still allowed from phenomenological
point of view. In this case, fermionic dark matter candidate $F$
annihilation in the mass region between $2~\GeV$ and $3~\TeV$
through $t$-channel exchange of $\eta$ can satisfy dark matter relic
density bound \cite{Vicente:2014wga}. And as a consequence of
hierarchal Yukawa structure, dark matter $F$ annihilates mainly into
third-family leptons: $\tau^+\tau^-$ and
$\nu_{\tau}\bar{\nu}_{\tau}$.

\subsection{Dark Matter}

In our model, a multi-component dark matter scenario is possible due
to the residual $Z_2'$ symmetry between two sets of new scalars and
fermions. For instance, we choose the lightest fermion $N_{R1}$
(refer as $F$) in set one and the lightest scalar $H_2^0$ (refer as
$S$) in set two as dark matter candidates. These dark matter
candidates must satisfy two experimental constrains : (1) the dark
matter relic density observed by Plank \cite{Ade:2013zuv}
$\Omega_{DM}h^2=0.1193\pm0.0014$; (2) cross section for direct
detection of dark matter scattering off nucleon set by LUX
\cite{Akerib:2013tjd}.

\begin{figure}[!htbp]
\begin{center}
\includegraphics[width=0.45\linewidth]{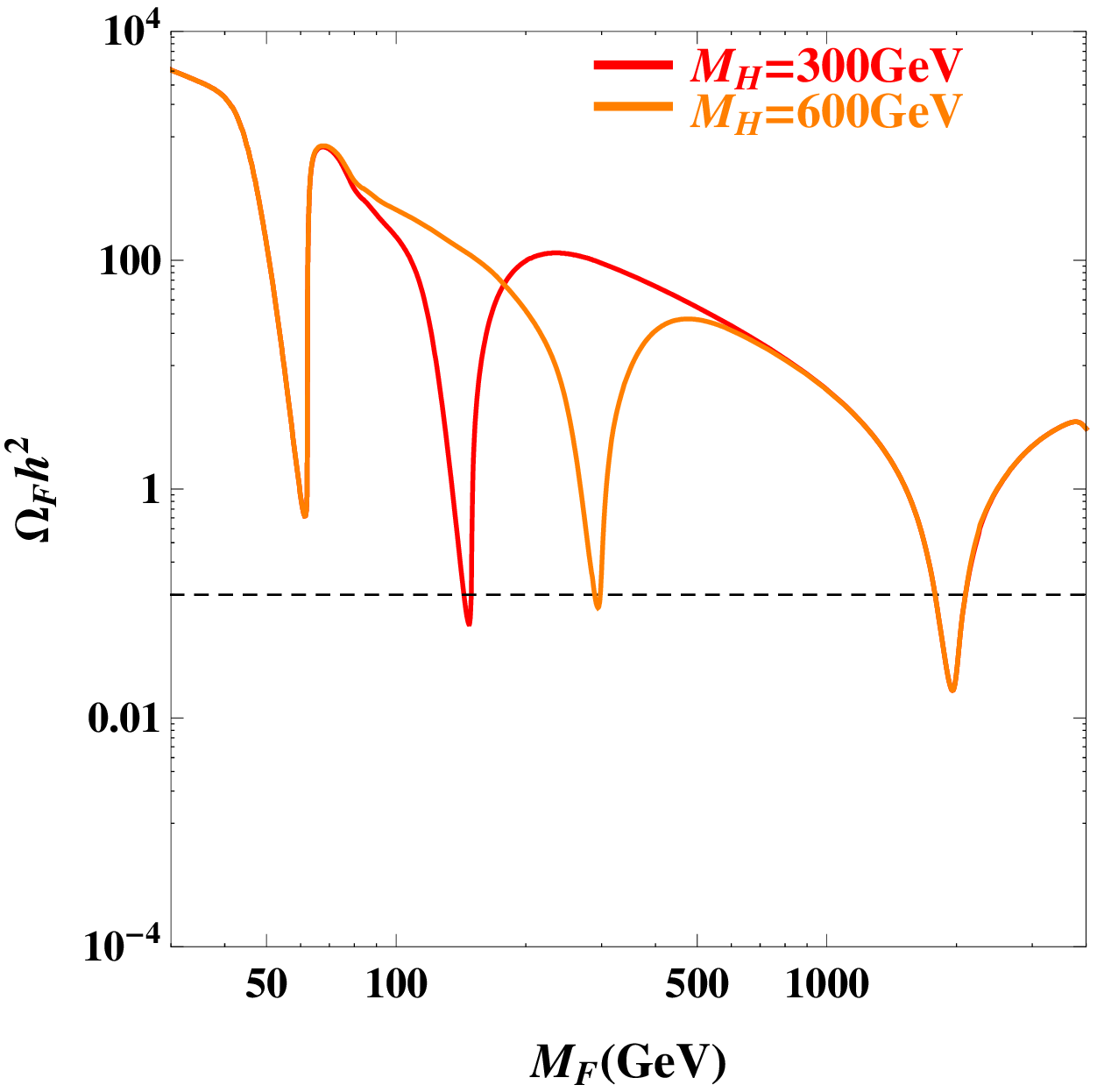}
\includegraphics[width=0.45\linewidth]{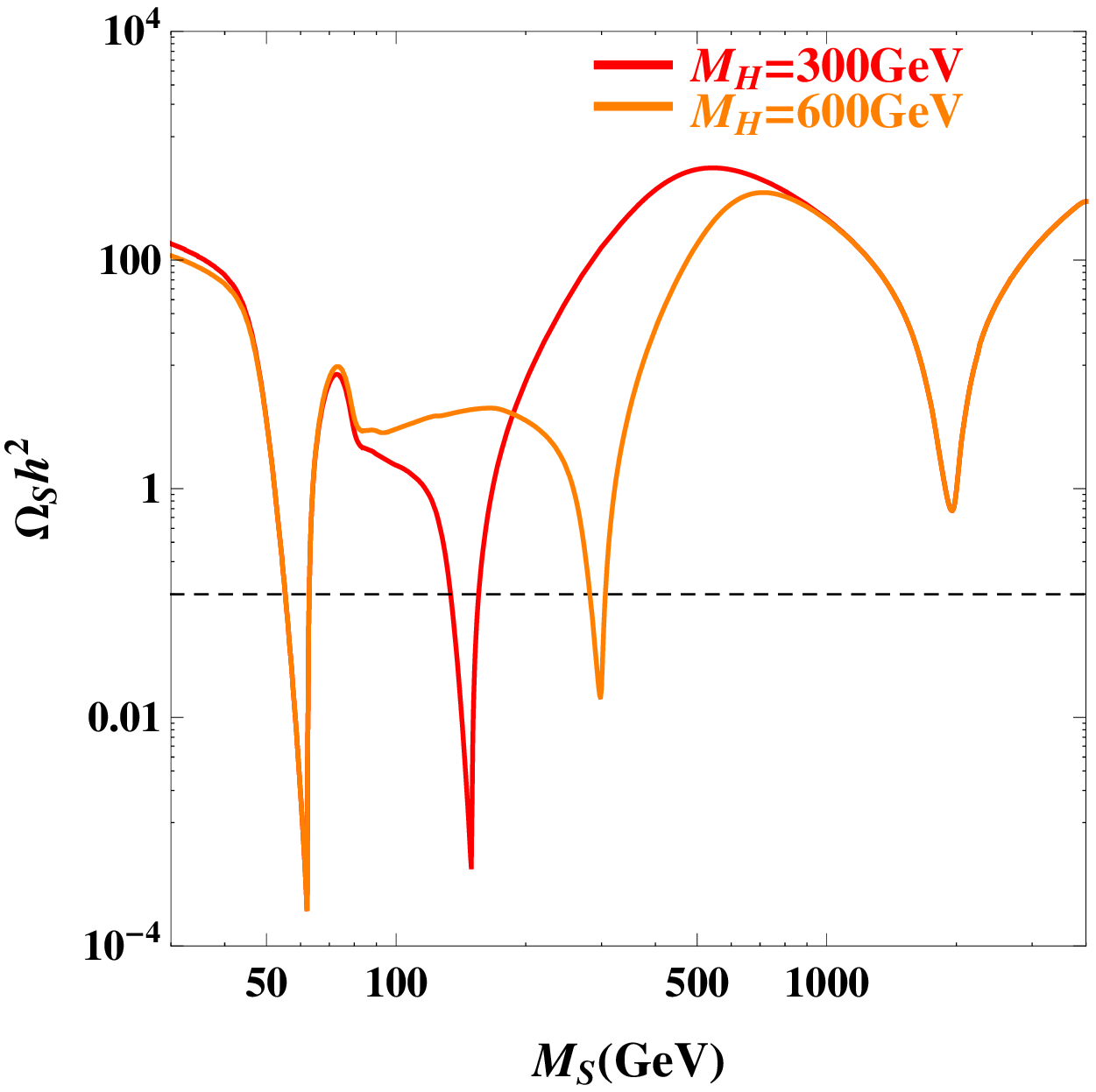}
\caption{Relic density of fermion (left) and scalar (right) dark
matter as a function of $M_{DM}$ for one dark matter candidate.
Here, we set the relevant scalar interaction coupling
$\lambda_{\eta_2\Phi}=\lambda^{\prime}_{\eta_2\Phi}=\lambda_{s_2\Phi}=-0.001$,
$\lambda_{\sigma\eta_2}=\lambda_{s_2\sigma}=-0.001$. We also fix
$M_h=125~\GeV$, $M_H=300/600~\GeV$, $\sin\theta=0.3$,
$M_{Z'}=4~\TeV$, and $v_{\sigma}=8~\TeV$. \label{RD}}
\end{center}
\end{figure}

Theoretical calculation of dark matter relic density is well
described in \cite{Bertone:2004pz}, and it is calculated with the help
of packages {\bf Feynrules} \cite{feynrules} and {\bf micrOMEGAs}
\cite{micromegas} in our analysis. Because the $t$-channel Yukawa
portal may suffer constrains from LFV or neutrino masses, and in our
benchmark point, the $t$-channel contribution to relic density is less
than 1\%, thus we will
focus on the $s$-channel $h/H/Z'$ portal for simplicity.
Firstly, the relic density of fermion/scalar dark matter
for one dark matter candidate is presented in Fig. \ref{RD},
where we neglect the conversion $F\bar{F}\leftrightarrow SS^*$
between the two dark matter candidates. For the fermion dark matter $F$,
the light $h$ portal can not acquire a sufficient annihilation cross
section, while the heavy $H$ portal is still promising when $M_F\sim
M_H/2$, which is because the suppression of large
$v_{\sigma}=8~\TeV$. Anyway, the $Z'$ portal can easily satisfy the
relic density when $M_F\sim M_{Z'}/2$. For the scalar dark matter
$S$, it is dominantly made from the singlet scalar $s_2^0$, due to
small mixing $\theta_2$. The relic density can easily be attained
when $M_S\sim M_h/2$ and $M_S\sim M_H/2$, while the $Z'$ portal is
not promising, mainly because the small $B-L$ charge of $S$ and
suppression of heavy $M_{Z'}$.

Secondly, we take into account the conversion of two component dark matter
$F\bar{F}\leftrightarrow SS^*$, which can be mediated by $s$-channel $h/H/Z'$,
where the $H$-portal is expected to be the dominant one.
Therefor, $HSS^*$ and $HF\bar{F}$ are the two most relevant couplings
to study conversion. For simplicity, we further assume
$\lambda_{\sigma\eta_2}=\lambda_{s_2\sigma}=\lambda$, which
determines $HSS^*$ coupling, and fix other parameters as discussed in
FIG. \ref{RD} if not mentioned.

\begin{figure}[!htbp]
\begin{center}
\includegraphics[width=0.45\linewidth]{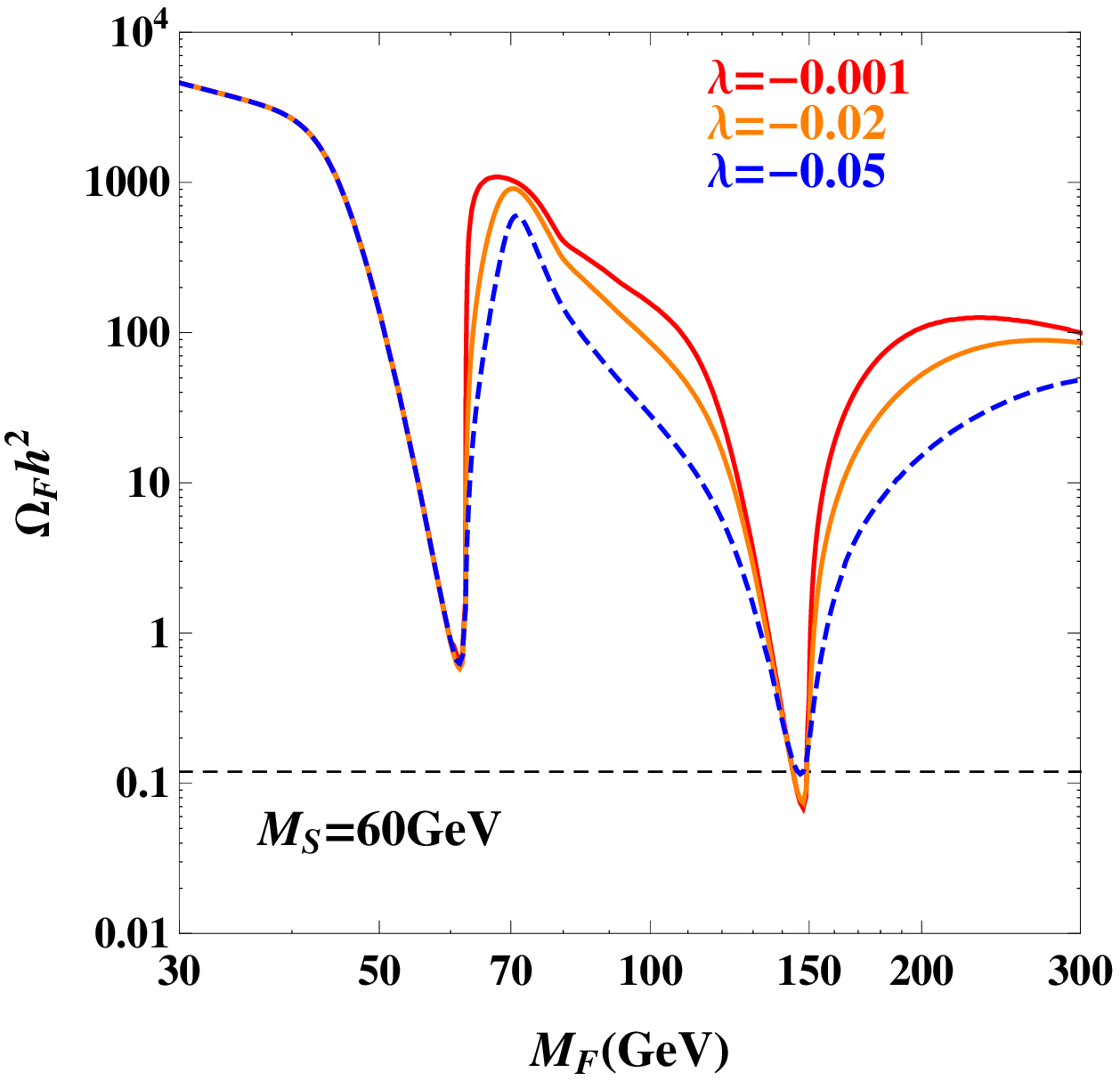}
\includegraphics[width=0.45\linewidth]{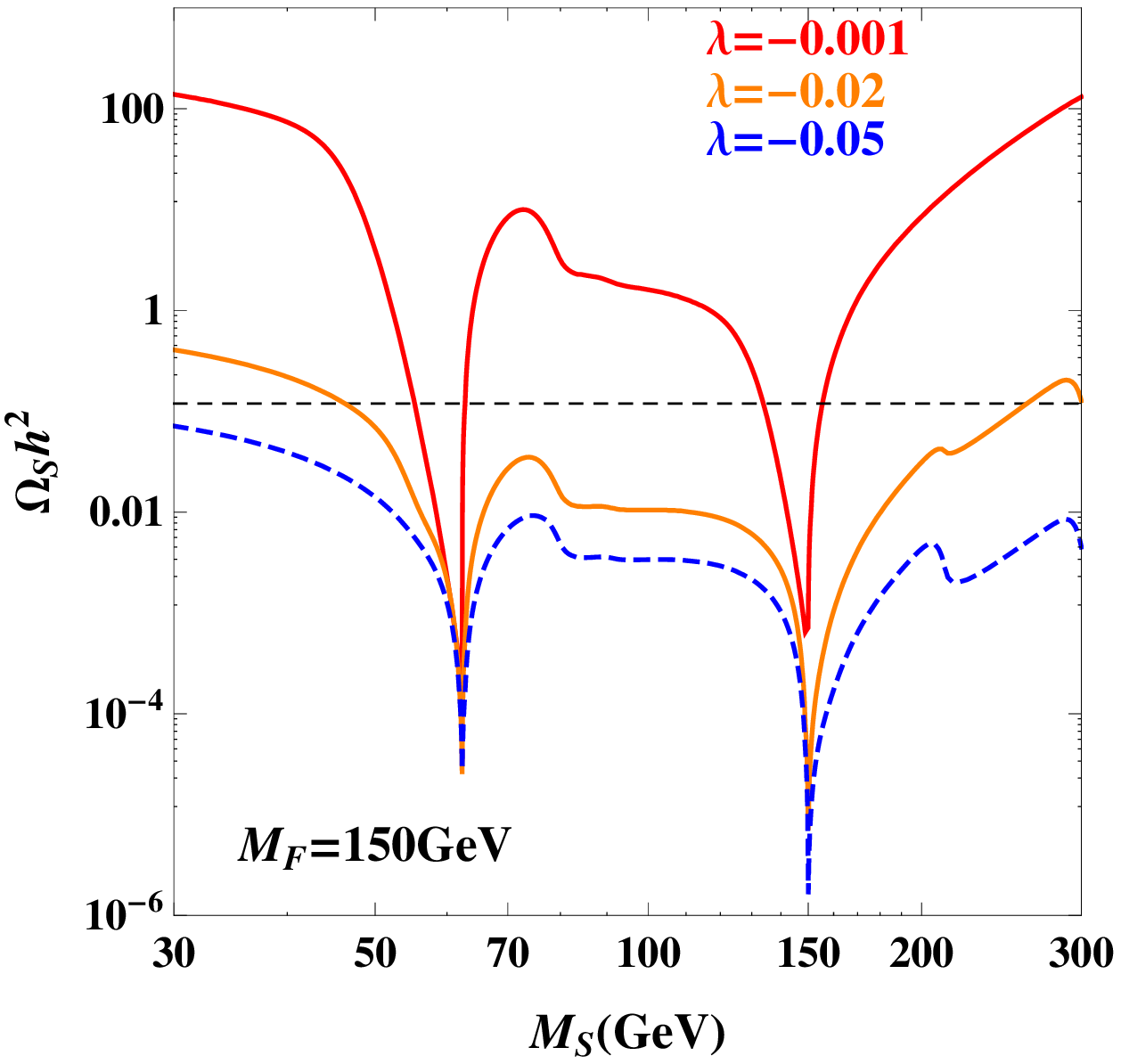}
\caption{ The dependence of $F/S$ relic density on $\lambda$.
\label{FSvlam}}
\end{center}
\end{figure}

The dependence of $F/S$ relic density on $\lambda$ is shown in FIG.
\ref{FSvlam}. For the fermion dark matter, when $M_F>M_S$, the
larger $\lambda$ the larger $F\bar{F}\leftrightarrow SS^*$
annihilation rate, and therefor the smaller the relic density. It is
clear that the $\Omega_F h^2$ can differ by about one order of
magnitude between $\lambda=-0.001$ and $\lambda=-0.05$. But when
$M_F<M_S$, the effect of conversion $F\bar{F}\to SS^*$ is quite
small. Noticing for $\lambda=-0.05$, the $H\to SS^*$ can greatly enhance
the total decay width of $H$, which causes the increase of $\Omega_F
h^2$ around $M_F\sim M_H/2$. For the scalar dark matter, the
increase of $\lambda$ will decrease the  relic density significantly
due to the increase of $SS^*$ annihilation. But, the friction of
conversion $SS^*\to F\bar{F}$ keeps the same, since the Yukawa
coupling $HF\bar{F}$ is fixed by $M_F$. The arguments are true if we
exchange the roles of $F$ and $S$, and fix $HSS^*$ coupling while
verify $HF\bar{F}$ coupling.

\begin{figure}[!htbp]
\begin{center}
\includegraphics[width=0.45\linewidth]{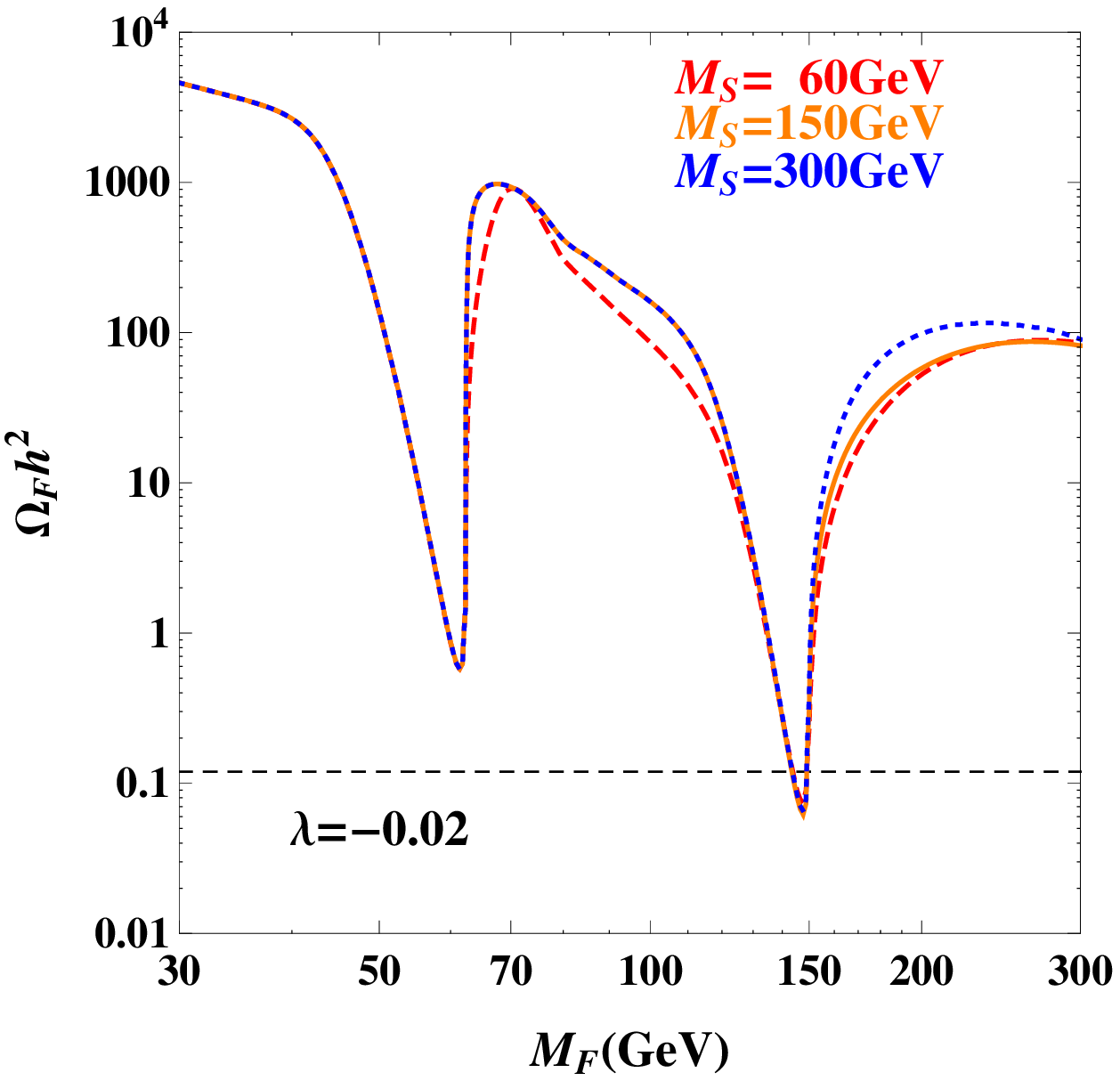}
\includegraphics[width=0.45\linewidth]{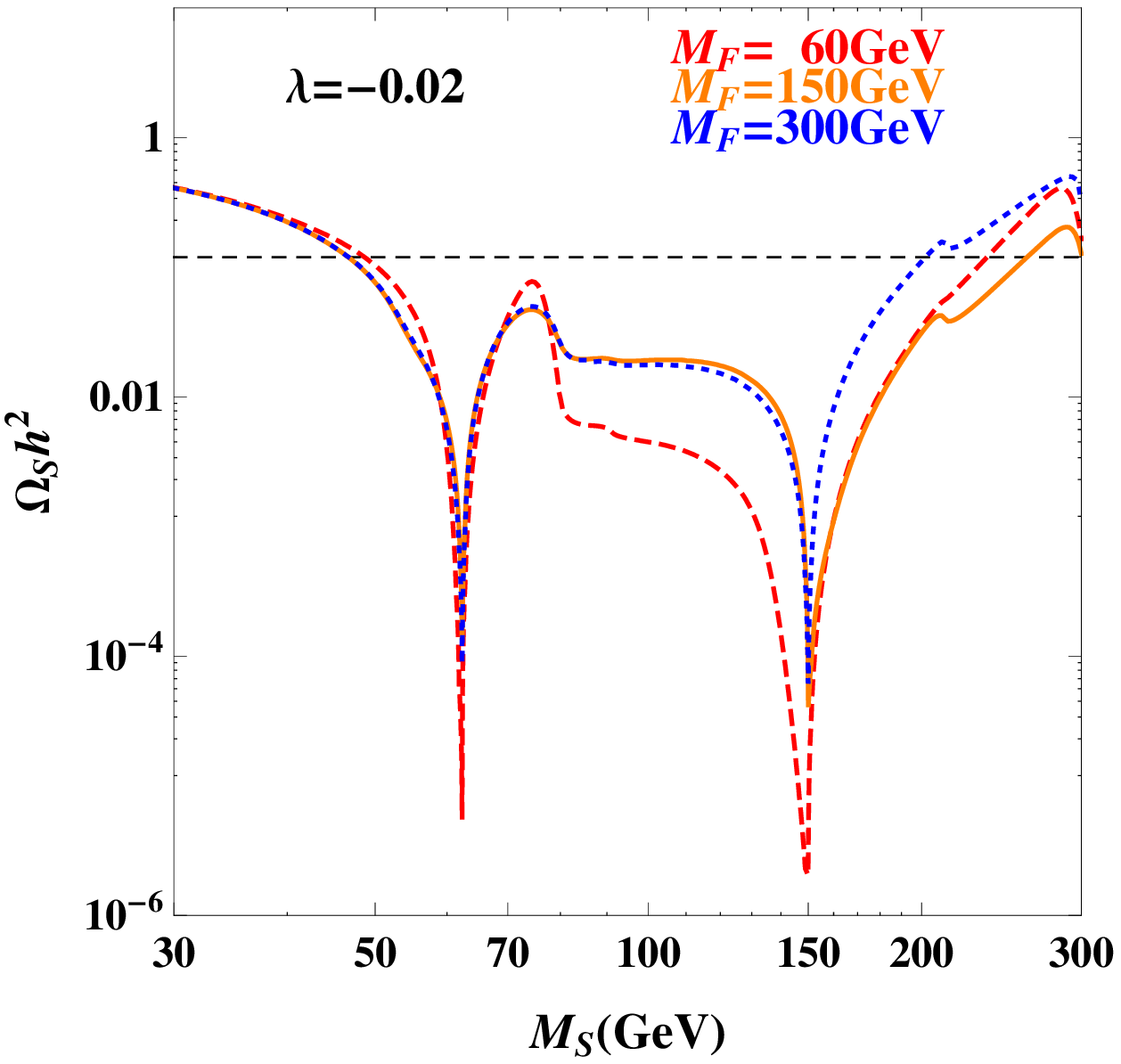}
\caption{ The effect of two component dark matter conversion on
$F/S$ relic density for fixed $\lambda=-0.02$. \label{FSvM}}
\end{center}
\end{figure}

Another aspect of conversion is the masses of the two dark matter.
The left(right) of FIG. \ref{FSvM} shows the $F(S)$ relic density
for $M_{S(F)}=60,150,300~\GeV$ with $\lambda=-0.02$. In case of
fermion dark matter, it is clear that the smaller the $M_S$ the
larger $F\bar{F}\rightarrow SS^*$ annihilation rate, and therefore
the smaller the relic density. For relative heavy
$M_S=150,300~\GeV$, the conversion has tiny effect on the $F$ relic
density when $M_F<M_H/2$. In case of scalar dark matter, the
dependence of $\Omega_Sh^2$ on $M_F$ is a little complicated, since
the $HF\bar{F}$ coupling is directly related with the $M_F$.
 For $M_S<80~\GeV$, the effect of conversion is relatively small,
 and one expect that the larger the $M_F$ the smaller
 the $S$ relic density, which is mainly caused by
the increase of $HF\bar{F}$ coupling. In medium mass region
$80<M_S<200~\GeV$, the conversion effect would be dominant, thus the
smaller the $M_F$ the smaller the $S$ relic density. In the high
mass region, the conversion effect is comparable to $HF\bar{F}$
coupling effect, which makes the dependence of $\Omega_Sh^2$ on
$M_F$ nonlinear. In a word, the $HSS^*$ and $HF\bar{F}$ couplings play
vital importance in dark matter conversion.
 The conversion can take place in both
direction $F\bar{F}\to SS^*$ and $SS^*\to F\bar{F}$ when $M_F\sim
M_S$, which can be obtained when both $F$ and $S$ are mainly
annihilation through $H$-portal. If not the case, only the
conversion of heavier one into light one is relevant
\cite{Esch:2014jpa}.

Finally, we discuss the constrains from direct detection of dark
matter. The Current experiment constraints assume the existence of
only one dark matter specie. However two-component dark matter
candidates are predicted in our model. Therefore the contribution of
cross section on nucleon for each specie should be rescaled by the
fraction factor of relic density. We define the fraction of the mass
density of $i$-th dark matter in case of multi-component dark matter
\cite{Cao:2007fy,Aoki:2012ub}:
\begin{equation}
\epsilon_i = \frac{\Omega_i h^2}{\Omega_{CDM} h^2},
\end{equation}
where $i=F,S$ in our consideration. Thereafter, the up limit of
direct detection is:
\begin{equation}\label{DD}
\frac{\epsilon_F}{M_F} \sigma_{F-N} + \frac{\epsilon_{S}}{M_{S}}
\sigma_{S-N} < \frac{\sigma_{exp}}{M_{DM}}
\end{equation}
Here, $\sigma_{F-N}(\sigma_{S-N})$ denotes the scattering cross
section of $F(S)$ with a nucleon $N$. The benchmark point in Eqn.
\ref{bp} gives the spin-independent scattering cross section
$\sigma_{F-N}^{SI}=1.10\times10^{-46}~\cm^2$
($\sigma_{S-N}^{SI}=1.62\times10^{-44}~\cm^2$) with $\Omega_F
h^2=1.12\times10^{-1}$ ($\Omega_S h^2=2.64\times10^{-4}$). Although
the bare $\sigma_{S-N}^{SI}$ is larger than the LUX upper constraint
$1.1\times10^{-47} \cm^2/\GeV$\cite{Akerib:2013tjd}, the
contribution of scalar $S$ to the scattering on nucleon is
suppressed because of its small faction
$\epsilon_S=2.21\times10^{-3}$. The value of expression on left of
Eqn. \ref{DD} is $1.3\times10^{-48} \cm^2/\GeV$, which is smaller
than current LUX bound. Thus, the fermion dark matter is dominant in
this scenario, while the scalar dark matter must be less than $4\%$
to escape current LUX bound.

\subsection{Collider Signatures}

As shown in our bench mark point (Eqn. \ref{bp}), the new particles
are all at the electroweak scale, which makes them testable at LHC.
Interactions between these new particles and SM Higgs $h$ will of
course modify the properties of $h$, thus give us some indirect hints.
Nowadays the most precise measurement of $M_h$ is the combined
results of ATLAS and CMS \cite{Aad:2015zhl}:
\begin{equation}
M_h=125.09\pm0.21(\mbox{stat.})\pm0.11(\mbox{sys.})~\GeV
\end{equation}
Apparently, the extra of new scalars and fermions would change
the decay rates of SM Higgs $h$. For instance, mixing between $h$ and
additional scalar singlet $H$ will modify tree-level $h$ decays.
And the additional charged-scalars $\eta^\pm_{1,2}$ will contribute
to the loop-induced decays as $h\to \gamma\gamma$ \cite{Arhrib:2012ia}.

It's well known that, for Higgs-portal dark matter, upper limits on Higgs
invisible decay is also interpreted an upper limits on dark
matter-nucleon scattering cross section \cite{Baek:2014jga}. Direct
measurement of Higgs invisible decay  in associated with $Z$ by
ATLAS set an upper limit of 75\% at 95\% C.L. \cite{Aad:2014iia}.
Combined analysis with Higgs signal strength give a more tight upper
limit of 37\% at 95\% C.L. \cite{Bechtle:2014ewa,Corbett:2015ksa,Aad:2015gba}.
In the future, the weak boson fusion channel might have the ability
to probe invisible decay to 2-3\% with $3000~\fb^{-1}$ at LHC
\cite{Bernaciak:2014pna}. Light dark matter candidates in our model will
contribute the the invisible decays of $h$ in our model.
In case of scalar dark matter, the branching ratio of Higgs invisible decay,
i.e., BR($h\to SS^*$) is $1.7\%$ for $M_{S}=62~\GeV$ with
$\lambda_{s_2\Phi}=\lambda_{s_2\sigma}=-0.001$, $\sin\theta=0.3$.
On the other hand for fermion dark matter, BR($h\to F\bar{F}$) is
$1.7\times10^{-6}$ for $M_F=62~\GeV$ with $v_{\sigma}=8~\TeV$,
$\sin\theta=0.3$, although it might not be favored by the constrains
on relic density of dark matter.

\begin{figure}[!htbp]
\begin{center}
\includegraphics[width=0.45\linewidth]{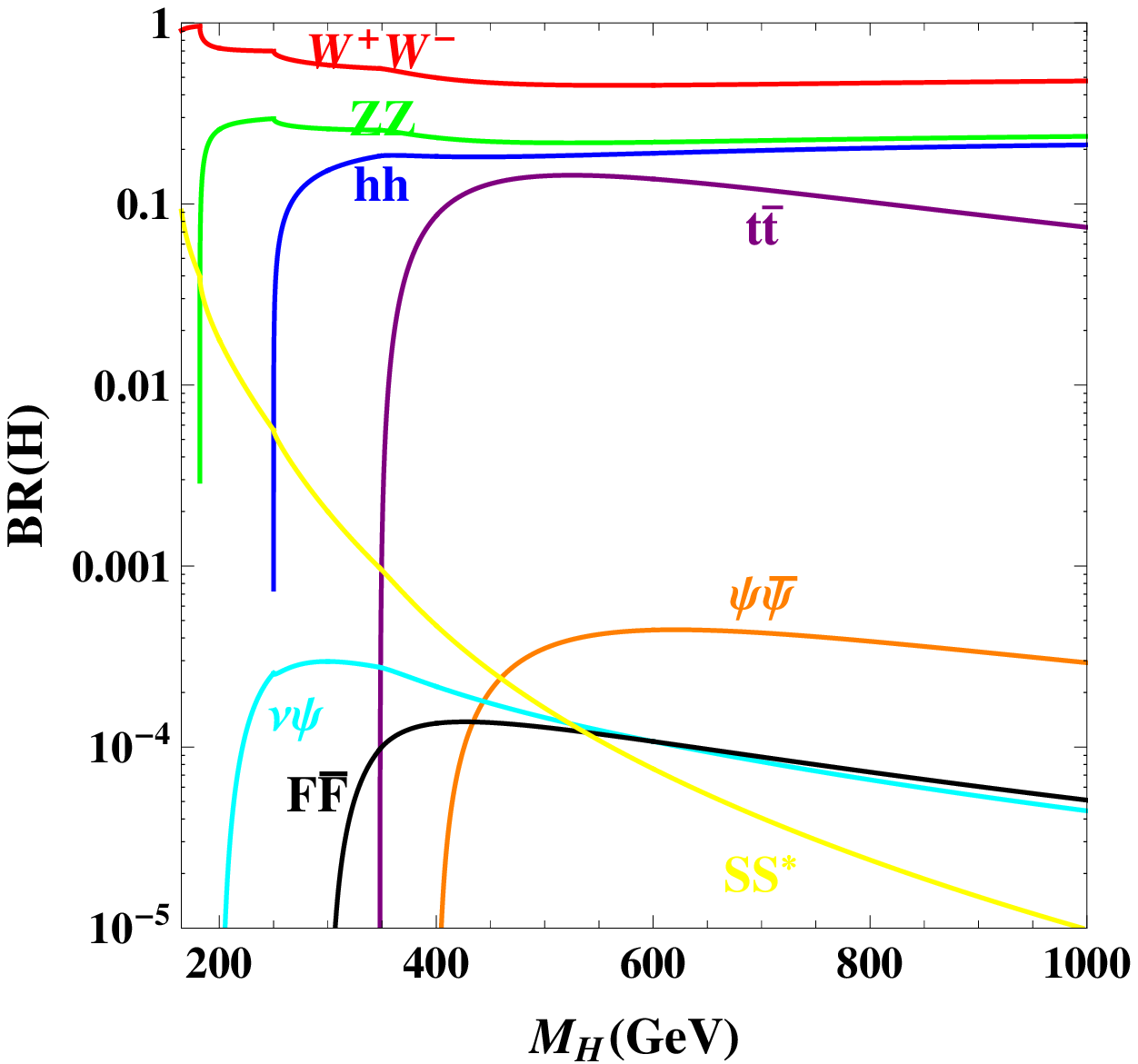}
\includegraphics[width=0.45\linewidth]{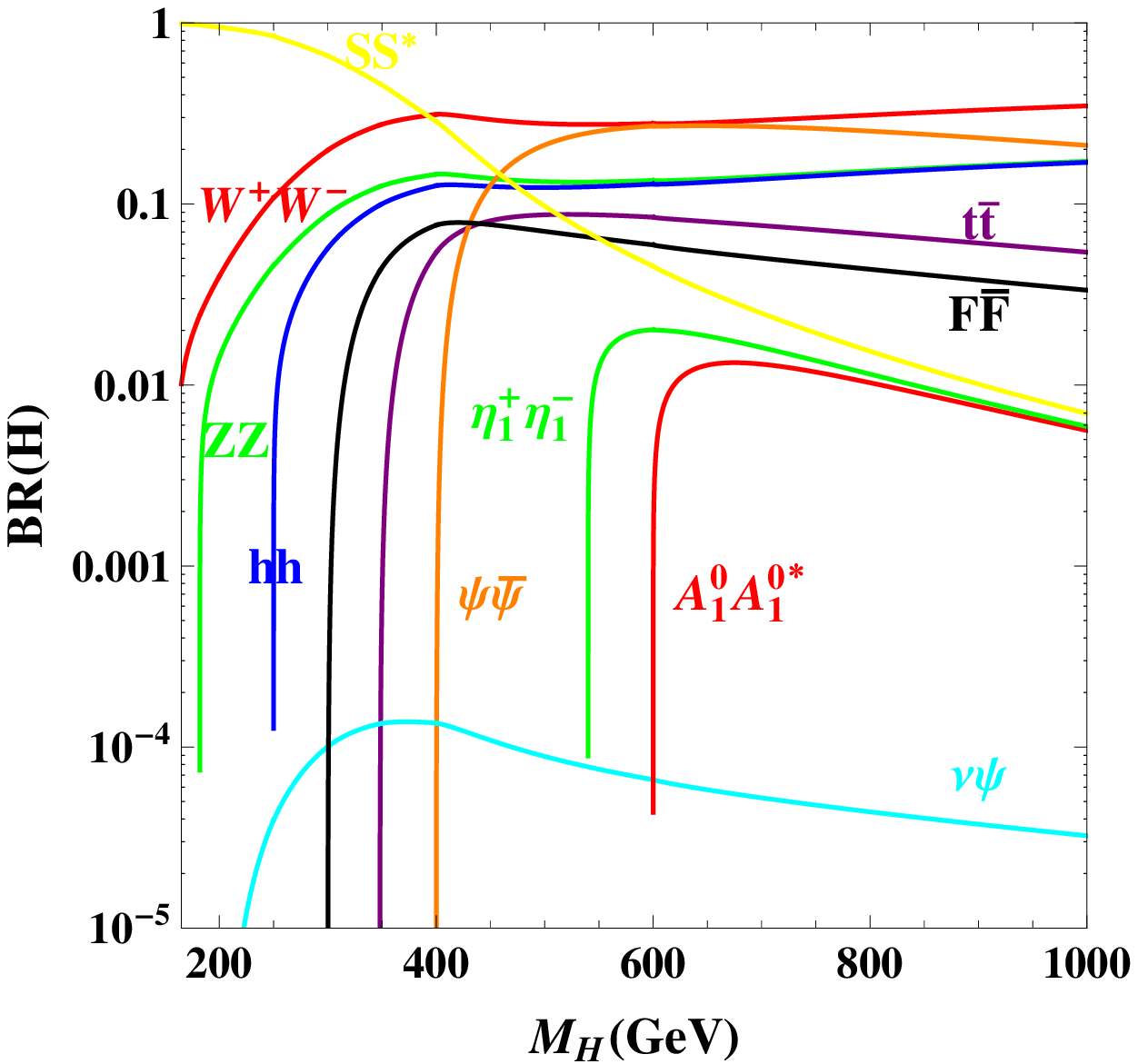}
\caption{Branching ratios of $H$ as a function of $M_H$ for
$\sin\theta=0.3$(left) and $\sin\theta=0.01$(right) in our benchmark
point in Eqn. \ref{bp}. \label{BRH}}
\end{center}
\end{figure}

Then we discuss the mixing between $h$ and $H$. The analysis of signal
strength of $h$ constrains $\sin^2\theta <0.23$ at 95\% C.L.
\cite{Giardino:2013bma,Falkowski:2013dza}. Direct search of $H$ in
the $ZZ$ and $WW$ channel now has push this limit down to
$\sin^2\theta <0.1$ with no new physics contribute to the decays of
$H$ \cite{Pelliccioni:2015hva}. Future hadron collider, i.e.,
HL-LHC, has the ability to probe $\sin^2\theta \sim4\times10^{-2}$,
and lepton collider, i.e., CEPC, could reach $\sin^2\theta
\sim2\times10^{-3}$ \cite{Dawson:2013bba}. In Fig. \ref{BRH}, we
show the branching ratios of $H$ for two values of $\sin\theta(0.3,0.01)$.
For a relatively large mixing angle $\sin\theta =0.3$, the
heavy neutral Higgs $H$ decays dominantly into SM particles. The
branching ratio of invisible decay $H\to SS^*$ can reach $10\%$ for
$M_H\sim165~\GeV$, and we expect it becomes dominant when
$M_H<160~\GeV$. The branching ratios of $H$ decaying into other new
physical particles are below $10^{-3}$ in this case. While for
$M_H\gg M_W$, it is well known that decays of $H$ into vector bosons
are determined by their Goldstone nature, which implies:
\begin{equation}
\mbox{BR}(H\to hh)\approx\mbox{BR}(H\to
ZZ)\approx\frac{1}{2}\mbox{BR}(H\to WW)
\end{equation}
The asymptotic behavior of this relation is clear shown on left
picture of Fig. \ref{BRH}. On the other aspect, for a relatively
small mixing angle $\sin\theta =0.01$, decays of $H$ into SM
particles will be suppressed and decays into new particles will be
greatly enhanced. $H\to SS^*$ is dominant when $M_H<400~\GeV$. The
branching ratio of $\psi\bar{\psi}$ will reach about 0.25 when
$M_H\sim 600~\GeV$, which is comparable with $H\to W^+W^-$. In this
case, $H\to F\bar{F}$ is below $10\%$ and $H\to
\eta^+\eta^-/A_1^0A_1^{0*}$ is below $2\%$.

The heavy neutral Higgs $H$ is testable for large mixing angle $\theta$.
For example, the promising channels to probe
heavy neutral Higgs $H$ would be $ZZ\to 4l$, $ZZ\to 2l2\nu$, $ZZ\to
2l2j$, $ZZ\to 2l2\tau$, $WW\to 2l2\nu$, $WW\to l\nu2j$, $hh\to 4b$,
and $hh\to 2b2\gamma$ \cite{Buttazzo:2015bka}. At the same time, we
would like to mention that the heavy Higgs $H$ could enhance the
di-Higgs production $hh$ \cite{Baglio:2012np} by a factor of 18
comparing to standard model case \cite{Chen:2014ask}. For small
mixing angle $\theta$, production of $H$ will be suppressed by this
small $\theta$, thus makes it challenging to probe directly at
colliders.

\begin{figure}[!htbp]
\begin{center}
\includegraphics[width=0.45\linewidth]{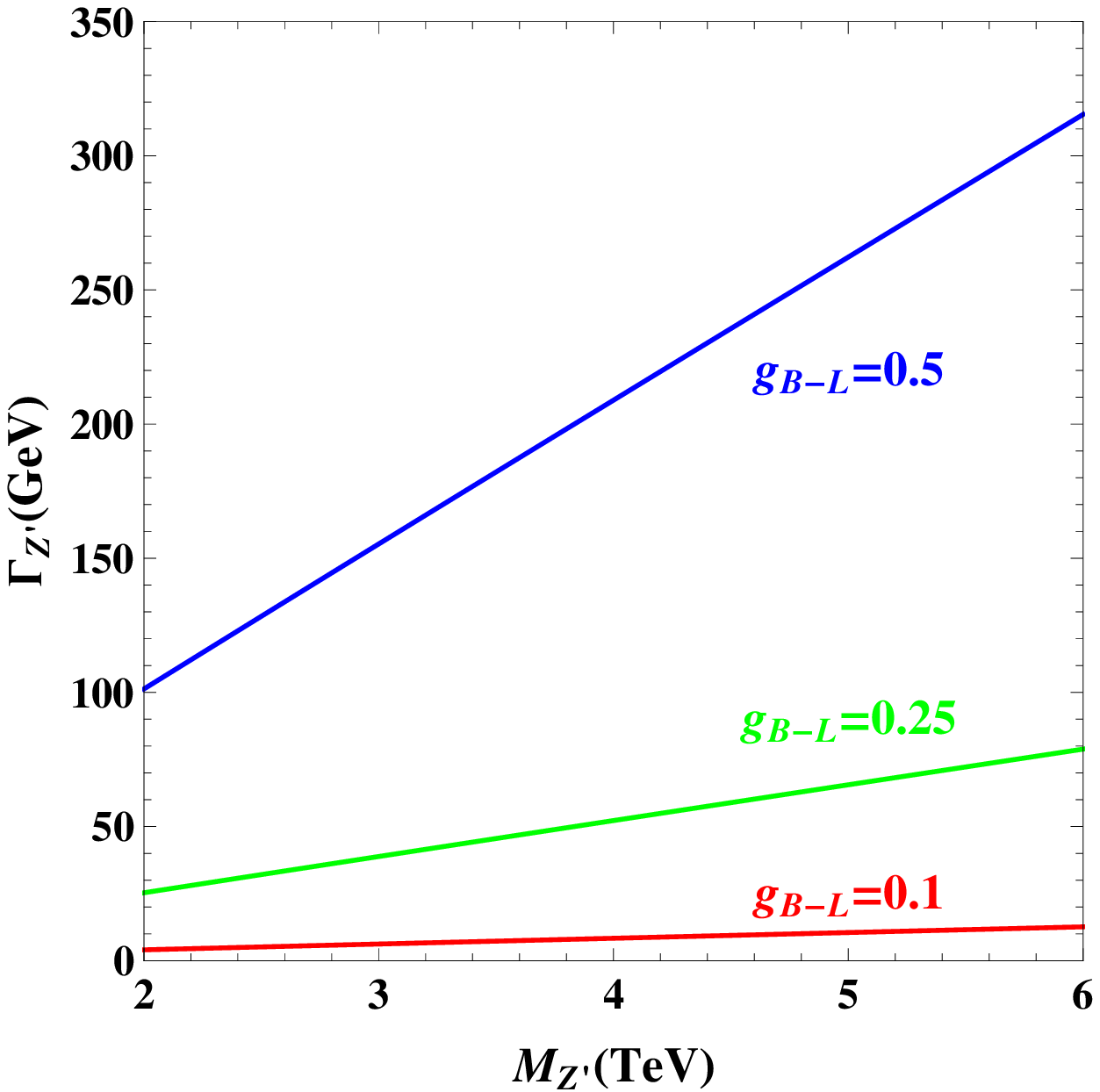}
\includegraphics[width=0.45\linewidth]{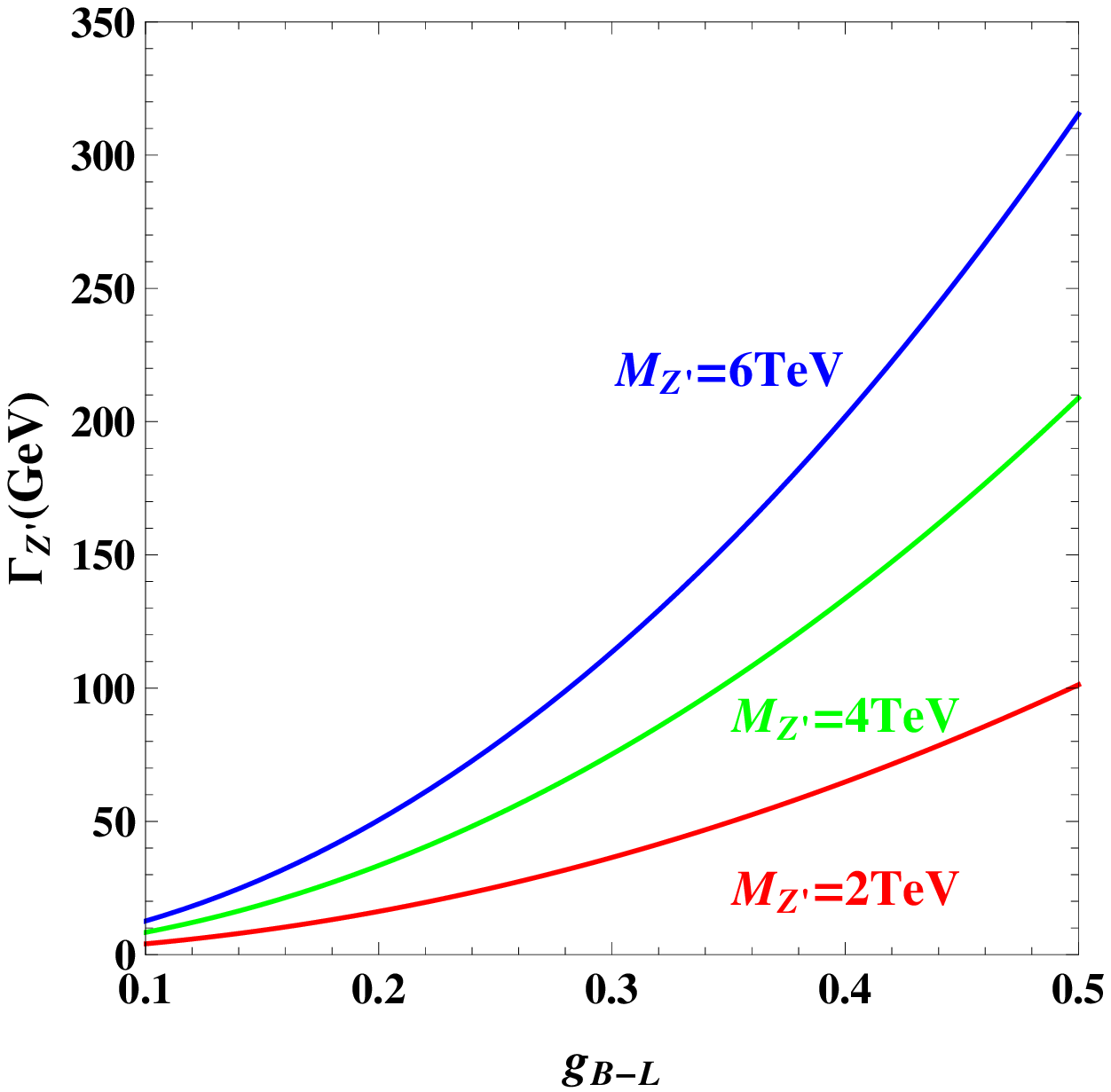}
\caption{Total decay width of $Z'$ as a function of $M_{Z'}$ (for
fixed values of $g_{B-L}$), and $g_{B-L}$ (for fixed values of
$M_{Z'}$). \label{ZWidth}}
\end{center}
\end{figure}

Next we review the properties of $U(1)_{B-L}$ gauge boson $Z^\prime$.
With about $20~\fb^{-1}$ data at $8~\TeV$ LHC,
bound on $Z'$ has been push up to $2.95~\TeV$ by CMS through the ratio
$R_{\sigma}=\sigma(pp\to Z'\to \ell^+\ell^-)/\sigma(pp\to Z\to \ell^+\ell^-)$,
where $\ell=e,\mu$ \cite{Khachatryan:2014fba}. In our benchmark point,
we choose $M_{Z'}=4~\TeV$ and $g_{B-L}=0.5$ ($v_{\sigma}=8\TeV$),
which can safely satisfy current experimental limits and can be tested
at $14~\TeV$ LHC with $100~\fb^{-1}$ \cite{Basso:2008iv,Basso:2010pe}.
Fig. \ref{ZWidth} shows the total decay width of $Z'$ as a function of
$M_{Z'}$ and $g_{B-L}$. Depending on $g_{B-L}$, $\Gamma_{Z'}$ varies from
a few to hundreds of GeV. For such large $\Gamma_{Z'}$, it can be directly
measured by the leptonic final states at LHC \cite{Basso:2008iv,Basso:2010pe}.

In table \ref{ZBranch}, we give the decay branching ratios of $Z'$
in our benchmark point. The dominant decay channels of $Z'$ are
$q\bar{q},l\bar{l}$, and $\nu_L\bar{\nu}_L$, while all of the new
particle final states only account for about $20\%$. A distinct
feature of $U(1)_{B-L}$ gauge boson $Z'$ is the definite relation
between quark and lepton final states:
 \begin{equation}
 \mbox{BR}(Z'\to q\bar{q}):\mbox{BR}(Z'\to l\bar{l}) \simeq 2 : 3
 \end{equation}
after summing over all flavors. This relation can be used to
distinguish $U(1)_{B-L}$ gauge boson $Z'$ from $Z'$ in other models
\cite{Langacker:2008yv}. More practical on experiment, the B-L
nature of $Z'$ can be tested if BR$(Z'\to b\bar{b})/$BR$(Z'\to
\mu^+\mu^-)=1/3$ is confirmed \cite{Kanemura:2011mw}. In our model
with only left-handed light neutrinos, the dominant invisible decay
channel of $Z'$ is BR$(Z' \to \nu_L\bar{\nu}_L)$, which is half of
BR$(Z' \to l\bar{l})$. Further with dark matter candidate in our
model, $Z'$ invisible decays get additional contributions from $Z'$
into dark matter pairs. For instance, BR$(Z'\to \mbox{inv.})$ could
be 0.2457, 0.1990, 0.1964 for $FF$, $FS$ and $SS$ dark matter
separately. So precise measurement of BR$(Z'\to \mbox{inv.})$ will
shed light on the nature of dark matter.

 \begin{table} [!htbp]
\begin{center}
\begin{tabular}{|c|c|c|c|c|c|c|c|c|c|}
\hline
$q\bar{q}$ & $l\bar{l}$ & $\nu_L\bar{\nu}_L$ & $\psi\bar{\psi}$  & $N_{R1}\bar{N}_{R1}$  & $N_{R2}\bar{N}_{R2}$ & $\chi_1\bar{\chi_1}$ & $\chi_2\bar{\chi_2}$  \\
\hline
  0.25       &    0.38      &     0.19      &  0.063  & 0.0077 & 0.0077 & 0.016 & 0.032 \\
\hline
$HH$ & $hh$ & $A^0_1 A_1^{0*}$ & $H^0_1 H_1^{0*}$  & $A^0_2 A_2^{0*}$  & $H^0_2 H_2^{0*}$ & $\eta_1^+\eta_1^-$ & $\eta_2^+\eta_2^-$  \\
\hline
   0.030      &   0       &   0.0076   &  0.0051   &  0.0010 & 0.0013 & 0.0076 & 0.0010 \\
\hline
\end{tabular}
\end{center}
\caption{Branching ratios of $Z'$ in our benchmark point. Here, we
set $M_H=300~\GeV$ and $\sin\theta=0$ for simplicity.}
\label{ZBranch}
\end{table}

 \begin{figure}[!htbp]
\begin{center}
\includegraphics[width=0.45\linewidth]{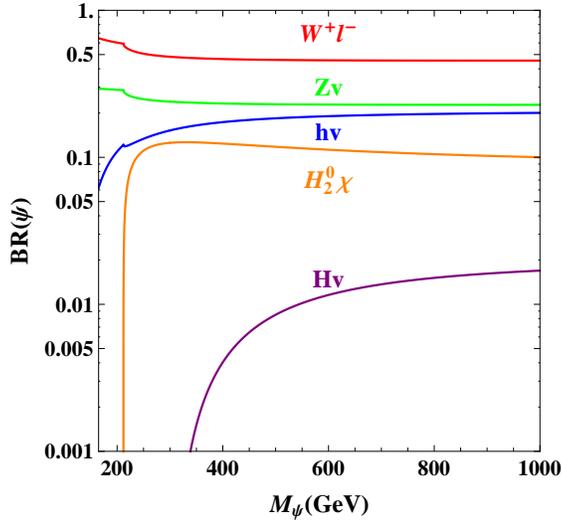}
\caption{ Branching ratios of $\psi$ as a function of $M_\psi$. We
have fixed $\sin\theta=0.3$, $M_H=300~\GeV$ and $y=h=0.01$.
\label{BRPsi}}
\end{center}
\end{figure}

Another interesting feature of our model is the existence of heavy Dirac fermion $\psi$,
thus there are no lepton number violation (LNV) decays as $\psi\to W^-l^+$.
For $M_{\psi}<M_h$, the Higgs decay into a pair of light and heavy neutrinos,
$h\to \bar{\nu} \psi+\bar{\psi}\nu$ will open, which could increase $\Gamma_h$
by up to almost 30\% and significantly affects Higgs searches at the LHC \cite{Chen:2010wn}.
In this paper, we consider $M_{\psi}>M_h$. Therefor, decay channels of $\psi$ could
be $W^+l^-$, $Z\nu$, $h\nu$, and if kinematically allowed $H\nu$, $A^0_iN_j$,
$A^0_i\chi_j$, $H^0_iN_j$, $H^0_i\chi_j(i,j=1,2)$ are also possible.
Due to tiny mixing angle $\theta_{1,2}$, branching ratios of
$\psi\to A^0_iN_j,A^0_i\chi_j$ are negligible. In Fig. \ref{BRPsi},
we show the branching ratios of $\psi$. It is clear that $\psi$ will decays
dominantly into standard model final states for comparable Yukawa couplings
of $y$ and $h$. Approximately for $M_{\psi}\gg M_W$, we have:
 \begin{equation}
\frac{1}{\cos^2\theta} \mbox{BR}(\psi\to h\nu)\approx
\mbox{BR}(\psi\to Z\nu)\approx\frac{1}{2} \mbox{BR}(\psi\to W^+l^-)
 \end{equation}
Decays of $\psi$ into new physical particles is small in this case.
BR$(\psi\to H^0_2\chi)$ is about $10\%$, once is kinematically opened.
BR$(\psi\to H\nu)$ is suppressed by $\sin^2\theta$, thus it is
always much smaller. As shown in table \ref{ZBranch},
BR$(Z'\to\psi\bar{\psi})\simeq0.063$ for one generation in our model,
so $\psi\bar{\psi}$ can be produced through $Z'$ portal.
A possible promising signature is the tri-lepton channel \cite{Basso:2008iv}:
 \begin{equation}
 pp\to Z' \to \psi \bar{\psi} \to W^+l^-+W^-l^+\to 2l^\pm l^\mp jj + \cancel{E}_T
 \end{equation}
The cross section of this tri-lepton signal is about 0.017 fb in our benchmark point,
so the tri-lepton is only promising at future high-luminosity LHC.
The mass of $\psi$ can be reconstructed using the transverse mass of
two opposite sign leptons with missing transverse momentum \cite{Basso:2008iv}.
Another feature of $\psi$ is the possible large mixing with $\nu_L$ comparing to
canonical type-I seesaw \cite{type1}. As discussed in Sec. \ref{neutrinomass},
the mixing $V_{\nu\psi}$ between $\nu_L$ and  $\psi_L$ is $M_D/\sqrt{M_D^2+M_\psi^2}$.
For $M_D\sim \mathcal{O}(1)\GeV$, $M_\psi\sim\mathcal{O}(100)\GeV$,
$V_{\nu\psi}\sim\mathcal{O}(10^{-2})$. Thereafter, $\psi$ could be
largely associated produced with charged leptons through $W$ \cite{Bambhaniya:2014kga}:
 \begin{eqnarray}
 pp\to W^* \to l^\pm\psi \to l^\pm + W^\pm l^\mp \to l^\pm l^\mp jj \\
 pp\to W^* \to l^\pm\psi \to l^\pm + W^\pm l^\mp  \to l^\pm l^\mp l^\pm \cancel{E}_T
 \end{eqnarray}
The production cross section $\sigma(l^\pm\psi)=350\times|V_{i\psi}|^2 ~\fb$ in our
benchmark point. And it might be promising at 14TeV LHC with about $100~\fb^{-1}$.
Testability of this heavy Dirac neutrino $\psi$ is less promising
than the heavy Majorana neutrinos with same mixing scale, since the
latter could rise LNV signatures \cite{Han:2006ip,del
Aguila:2007em,Atre:2009rg,Deppisch:2015qwa}.

Finally, we discuss the decays of inert scalars and fermions.
$N_{R1}$ and $H^0_2$ are a dark matter candidate as in our benchmark
point in Eqn. \ref{bp}. Decays of $N_{R2}$ are dominated by
$N_{R2}\to l^{\pm}\eta_1^{\mp*}\to l^{\pm}l^{\mp} N_{R1}$ and
$N_{R2}\to \nu A_1^{0*}\to \nu\nu N_{R1}$ through the Yukawa
coupling $f$. Decays of $\chi_i$ are $\chi_i\to H^0_2 \psi^*$ with
the off-shell $\psi^*$ further decaying into
$W^+l^-/Z\nu/h\nu/H^0_2\nu$. $A_1^0$ and $\eta_1^{\pm}$ mainly decay
through the Yukawa coupling $f$, which leads to $A_1^0\to\nu N_{Ri}$
and $\eta_1^\pm\to l^\pm N_{Ri}$. The heavy $Z_2$ odd scalar $H_1^0$
decays into $\psi N_{Ri}$ through Yukawa coupling $h_\alpha$ and
into $hA_1^0$ through trilinear coupling $\mu_1$. Similar, decays of
$A^0_2$ are $A^0_2\to \nu \chi_i$ and $A^0_2\to h H_2^0$, while
decays of $\eta_2^\pm$ are $\eta_2^\pm\to l^\pm \chi_i$ and
$\eta_2^\pm\to W^\pm H^0_2$.

\begin{table}[!htbp] \large
\begin{tabular}{|c|c|c|c|c|c|c|}\hline
Particles & $\eta_1^+\eta_1^-$ & $\eta^\pm_1 A_1^{0}$ & $A_1^0 A_1^{0*}$
          & $\eta_2^+\eta_2^-$ & $\eta^\pm_2 A_2^{0}$ & $A_2^0 A_2^{0*}$\\ \hline
$\sigma$ (in fb) & 5.8 & 16 & 3.6 & 0.089 & 0.31 & 0.075\\
\hline
\end{tabular}
\caption{Production cross sections for inert scalar doublets.}
\label{tabcs}
\end{table}

The inert scalar doublets can be pair production through DY process.
In Table \ref{tabcs}, we list the production cross sections for inert scalar doublets.
Many signatures can be risen from the inert particles.
In Ma's scotogenic model \cite{Ma:2006km},
promising signals of the doublet scalar on collider
are multi-lepton final states with missing transverse energy
 $\cancel{E}_T$ \cite{Dolle:2009ft,Miao:2010rg,Gustafsson:2012aj}.
 Similar signals can also be produced in our model, for example:
 \begin{eqnarray}
 2l + \cancel{E}_T & : & \eta^+_1\eta^-_1\to l^+ N_{R1}+l^- N_{R1} \\ \nonumber
                   & : & \eta^+_2\eta^-_2\to W^+ H^{0*}_2 + W^- H^0_2 \to l^+ \nu_l H^{0*}_2 + l^- \bar{\nu}_l H^0_2\\
 3l + \cancel{E}_T & : & \eta^+_1A^0_1 \to l^+ N_{R2}+\nu N_{R1}\to l^+ l^\pm l^\mp N_{R1}+\nu N_{R1} \\\nonumber
                   & : & \eta^\pm_2 A^0_2 \to W^\pm H^{0(*)}_2 + h H^0_2 \to l^\pm \nu_l H^{0(*)}_2 + l^+\nu_l l^- \bar{\nu}_l H^0_2 \\
 4l + \cancel{E}_T & : & \eta^+_1\eta_1^- \to l^+ N_{R2}+ l^- N_{R1}\to l^+ l^\pm l^\mp N_{R1}+ l^- N_{R1} \\\nonumber
                   & : & A^0_2 H^{0*}_2 \to h H^0_2 H^{0*}_2 \to l^+l^-l^+l^- H^0_2 + H^{0*}_2
 \end{eqnarray}
With much different decay topologies between our and Ma's model, it
would be distinguishable even with same signals. Apart from these
multi-lepton signals, there are also some other interesting signals
in our model, i.e.:
 \begin{eqnarray}
 2l^\pm jj + \cancel{E}_T & : & \eta^\pm_2 A^0_2 \to W^\pm H^{0(*)}_2 + h H^0_2 \to l^\pm \nu_l H^{0(*)}_2 + l^\pm \nu_l jj H^0_2\\
 l^\pm b\bar{b} + \cancel{E}_T & : & \eta^\pm_2 A^0_2 \to W^\pm H^{0(*)}_2 + h H^0_2 \to l^\pm \nu_l H^{0(*)}_2 + b\bar{b} H^0_2
 \end{eqnarray}
 The lepton number violation signal $2l^\pm jj + \cancel{E}_T$
 suffers much lower SM background, thus might make this signal
 very promising on LHC. The $l^\pm b\bar{b} + \cancel{E}_T$ has
 a relatively large production rate due to $h\to b\bar{b}$
 dominant in $h$ decay, so it might also be promising.

\newpage
\section{Conclusions}
\label{conclu} In usual canonical seesaw mechanisms, it requires the
heavy states with the scale of masses being grand unification scale
to generated the small neutrino mass. In linear seesaw scenario with
$m_{\nu}\simeq \mu_{L}M_{D}/M_{\Psi}$, the neutrino masses suffers a
two fold suppression by both lepton number symmetry violating term
$\mu_{L}$ and heavy mass $M_{\Psi}$. The linear seesaw model can
lower the seesaw scale such that new physics may arise at TeV scale.
In this work, we construct a radiated linear seesaw model where the
naturally small term $\mu_{L}$ are generated at one-loop level and
its soft-breaking of lepton number symmetry attributes to the SSB of
B-L symmetry at TeV scale. To satisfy the anomalies cancelation, the
value of B-L charges for inert particles are found to be exotic such
that there exists residual $Z_{2}\times Z_{2}^{\prime}$ symmetry
even after SSB of $B-L$ gauge symmetry. It is shown that the
residual symmetry stabilizes the inert particles as dark matter
candidates. In our model, we introduce two no-interplay classes of
inert particles to realize the model such that the lightest inert
particles belonging in each class play is the dark matter matter
candidate. Therefore we have propose a two-component dark matter
model. The seesaw scale of radiated linear seesaw scale can be as
low as a few hundred GeV, leading to interesting phenomenology.

Given a benchmark point at electro-weak scale, we illustrate the
main prediction of our model. For the Yukawa coupling
$f_{li}(f_{l})$ at 0.01 order, our benchmark point predicts
Br$(\mu\to e \gamma)\sim10^{-13}$, an order slightly under the
current constraints and in the reach of the forthcoming experiment.
The two component dark matter candidates are realized in our model.
To account for the observed relic density, the annihilation of dark
matter are dominant by the $s$-channel scalars $h/H$ or gauge boson
$Z^{\prime}$. For the fermion DM, we find that the $h$-channel is
excluded. But it is still allowed for the $H$-channel and
$Z^{\prime}$-channel. On the contrary, for the scalar DM, the
$Z^{\prime}$-channel is excluded while the $h/H$-channel is allowed.
And the heavy Higgs $H$ also plays a vital import rule in the
conversion between fermion and scalar dark matter. Collider
signatures of our model are also very rich. The precise measurements
of SM Higgs $h$ will put tight constrain light scalar DM and heavy
scalar $H$. With a relatively large mixing angle $\sin\theta=0.3$,
the $H\to ZZ,W^+W^-$ channels are testable at LHC. For the extra
$Z¡¯$ boson and heavy lepton $\psi$, the tri-lepton channel of
$Z¡¯\to\bar{\psi}\psi$ is promising at HL-LHC. With a larger cross
section, the associated production of $l^\pm\psi$ may be more
promising. The inert doublet scalar can also produce multi-lepton
channels. And some distinct channels, as $2\ell^\pm
jj+\cancel{E}_T$, $\ell^\pm bb+\cancel{E}_T$, can be used to
distinguish our model.

Finally, we would like to mention that the radiated linear seesaw
model we proposed is the minimal version where only one $\Psi$
fermion mediator is included. In this scenario, $M_{\nu}$ is a
rank-2 mass matrix and the lightest neutrino must be massless.
However, more complicated scenarios exist, corresponding to other
solutions of the anomaly free condition. Then one may obtain the
rank-3 neutrino mass matrix. Such scenarios predicts more new
particles with different $B-L$ charges, the model construction and
the phenomenology deserve us further study.

\section*{Acknowledgments}
We would like to thank Ran Ding for the help on the analysis of dark
matter. The work of Weijian Wang is supported by Special Fund of
Theoretical Physics, Grant No. 11447117 and Fundamental Research
Funds for the Central Universities.


\end{document}